\documentclass[prd,aps,showpacs,nofootinbib]{revtex4}
\usepackage{amssymb}
\usepackage{graphicx}
\usepackage{amsmath}

\newcommand{\be}{\begin{equation}}
\newcommand{\ee}{\end{equation}}
\newcommand{\bq}{\begin{eqnarray}}
\newcommand{\eq}{\end{eqnarray}}

\begin{document}\title{A fundamental explanation for the tiny value of the cosmological constant}
\author{Cl\'audio Nassif}
\affiliation{\small{CBPF-Centro Brasileiro de Pesquisas F\'{\i}sicas, Rua Dr.Xavier Sigaud,150,
CEP 22290-180, Rio de Janeiro-RJ, Brazil.}}

\begin{abstract}
We will look for an implementation of new symmetries in the space-time structure and their cosmological implications.
This search will allow us to find a unified vision for electrodynamics and gravitation. We will attempt to develop a
heuristic model of the electromagnetic nature of the electron, so that the influence of the gravitational field on the
electrodynamics at very large distances leads to a reformulation of our comprehension of the space-time structure
at quantum level through the elimination of the classical idea of rest. This will lead us to a modification of the
relativistic theory by introducing the idea about a universal minimum limit of speed in the space-time. Such a limit,
unattainable by the particles, represents a preferred frame associated with a universal background
field (a vacuum energy), enabling a fundamental understanding of the quantum uncertainties. The
structure of space-time becomes extended due to such a vacuum energy density, which leads to a negative pressure at
the cosmological scales as an anti-gravity, playing the role of the cosmological constant. The tiny values of the
vacuum energy density and the cosmological constant will be successfully obtained, being in agreement with current
observational results.
\end{abstract}

\maketitle

\section{Introduction}

   Motivated by Einstein's ideas for searching for new fundamental symmetries in
 Nature, our main focus is to go back to that point of the old
 incompatibility between mechanics and electrodynamics,by extending his
 reasoning in order to look for new symmetries that implement gravitation into
 electrodynamics of moving particles. We introduce more symmetries into the space-time
 geometry,where gravitation and electromagnetism become coupled with each
 other,in such a way to enable us to build a new dynamics that is compatible with the
quantum indeterminations.

 Besides quantum gravity at the Planck length scale,our new symmetry
 idea appears due to the indispensable
presence of gravity at quantum level for particles with very large
 wavelengths (very low energies).
 This leads us to postulate a universal minimum speed related to a
 fundamental (privileged)
reference frame of background field that breaks Lorentz
 symmetry\cite{1}.

 Similarly to Einstein's reasoning,which has solved that old
 incompatibility
 between nature of light and motion of matter (massive
 objects),
 let us now expand it by making the following heuristic assumption
 based on new
 symmetry arguments:

 {\it If,in order to preserve the symmetry (covariance) of Maxwell's
 equations,
 $c$ is required to be constant based on Einstein's reasoning,according
 to which it is impossible to find the rest reference frame for the speed
 of light ($c-c\neq 0$ ($=c$)) due to the coexistence of $\vec E$ and $\vec B$
 in
 equal-footing,then now let us think that fields $\vec E$ and $\vec B$
 may
also coexist for moving charged massive particles (as electrons),which
are at subluminal level ($v<c$). So,by making such an assumption,it
 would be also
 impossible to find a rest reference frame for a charged massive
 particle,by
 canceling
 its magnetic field,i.e.,$\vec B=0$ with $\vec E\neq 0$. This would
 break the coexistence of these two fields,which would not be possible
 because it is impossible to find a reference frame where
 $v=0$,in such a space-time. Thus we always must have $\vec E\neq 0$ and also $\vec B\neq 0$
for charged massive particles,due always to
the presence of a non-null momentum for the electron,in a similar way to the photon electromagnetic wave.}

 The reasoning above leads to the following conclusion:

  -{\it The plane wave for free electron is an idealization impossible to conceive
 under physical reality. In the event of an
 idealized plane wave,it would be possible to find the reference frame
 that cancels its momentum ($p=0$),just the same way as we can find the
 reference
frame of rest for classical (macroscopic) objects with uniform
 rectilineal motion (a state of equilibrium). In such an
 idealized case,
 we could find a reference frame where $\vec B=0$ for charged particle.
 However,
 the presence of gravity in quantum world emerges in order to always
 preserve
 the coexistence of $\vec E$ and $\vec B(\neq 0)$ in electrodynamics of
 moving massive particles (section 3). That is the reason why we think about a
 lowest and unattainable speed limit $V$ in such a space-time,
 in order to avoid to think about $\vec B=0$ ($v=0$). This means
 that there is no state of perfect equilibrium (plane wave and Galilean inertial
 reference frame) for moving particles in such a space-time,except the
 privileged inertial reference frame of a universal background field associated
 with an unattainable
 minimum limit of speed $V$. Such a reasoning allows us to think that
 the electromagnetic
 radiation (photon:$``c-c''=c$) as well as the matter (electron:
 $``v-v''>V(\neq 0)$) are in
 equal-footing,since now it is not possible to find a reference frame
 in equilibrium ($v_{relative}=0$) for both through any
 velocity transformations} (section 6).

  The interval of velocity with two limits $V<v\leq c$ represents the
 fundamental symmetry that is
 inherent to such a space-time,where gravitation and electrodynamics
 become coupled.
  However,for classical (macroscopic) objects,the breaking of that
 symmetry,i.e.,$V\rightarrow 0$,
 occurs so as to reinstate Special Relativity (SR) as a particular
 (classical) case,namely no uncertainties and no vacuum energy,where the idea
 of rest,based on the Galilean concept of reference frame is thus recovered.

  In another paper,we will study the dynamics of particles in the
 presence of
 such a universal (privileged) background reference frame associated
 with $V$,
 within a context of the ideas of Mach\cite{2},Schroedinger\cite{3}
 and
 Sciama\cite{4},where we will think about an absolute background
 reference frame in relation
 to which we have the inertia of all moving particles. However,we must
 emphasize that the approach
 we will intend to use is not classical as the machian ideas (the
 inertial reference
 frame of fixed stars),since the lowest limit of speed $V$,related to
 the privileged reference frame connected to a vacuum energy,has origin essentially
 from the presence of gravity at quantum level for particles with very large
 wavelengths.

   We hope that a direct relationship should exist between the minimum
 speed $V$ and Planck's minimum length $l_p=(G\hbar/c^3)^{1/2}(\sim 10^{-35}m)$
treated by Double Special Relativity theory (DSR)[20-25] (4th section).

 In the next section,a heuristic model will be developed to describe the
 electromagnetic
 nature of the matter. It is based on the Maxwell theory used for
 investigating the
 electromagnetic nature of a photon when the amplitudes of
 electromagnetic wave
 fields are normalized for one single photon with energy $\hbar w$.
  Thus,due to the reciprocity and symmetry reasoning,we shall extend such a
 concept for the matter (electron) through the idea of pair materialization after
$\gamma$-photon decay,so that we will attempt to develop a simple
 heuristic model of the electromagnetic nature of the electron that will experiment
a background field in the presence of gravity.

The structure of space-time becomes extended due to the presence of a vacuum energy density
associated with such a universal background field (a privileged reference frame connected to a zero-point
energy of background field,which is associated with the minimum limit of speed $V$ for particles moving
with respect to such a background reference frame). This leads to a negative pressure at the
cosmological length scales,behaving like a cosmological anti-gravity for the cosmological
constant whose tiny value will be determined (section 8).

\section{Electromagnetic Nature of the Photon and of the Matter}

\subsection{Electromagnetic nature of the photon}

In accordance with some laws of Quantum Electrodynamics\cite{5},we
shall take into account the electric field of a plane electromagnetic wave whose
amplitude is normalized for just one single photon\cite{5}. To do this,consider
that the vector potential of a plane electromagnetic wave is

\begin{equation}
\vec A=a cos(wt-\vec k.\vec r)\vec e,
\end{equation}
where $\vec k.\vec r=kz$,admitting that the wave propagates in the
 direction of z,being $\vec e$ the unitary vector of polarization.
 Since we are in vacuum,we must consider

\begin{equation}
\vec E=-\frac{1}{c}\frac{\partial\vec
 A}{\partial{t}}=(\frac{wa}{c})sen(wt-kz)
\vec e
\end{equation}

 In the Gaussian system of units,we have $|\vec E|=|\vec B|$.
 So the average energy density of the wave shall be

\begin{equation}
\left<\rho_{eletromag}\right>=
\frac{1}{8\pi}\left<|\vec E|^2+|\vec B|^2\right>=
\frac{1}{4\pi}\left<|\vec E|^2\right>
\end{equation}

 Substituting (2) into (3),we obtain

\begin{equation}
\left<\rho_{eletromag}\right>=
\frac{1}{8\pi}\frac{w^2a^2}{c^2},
\end{equation}
where $a$ is an amplitude that depends upon the number of photons in
 such a wave. Since we wish to obtain the plane wave of one single
 photon
 ($\hbar w$),then by making this condition for (4) and by considering an
 unitary
 volume for the photon ($v_{ph}=1$),we have

\begin{equation}
a=\sqrt{\frac{8\pi\hbar c^2}{w}} 
\end{equation}

 Substituting (5) into (2),we obtain

\begin{equation}
\vec{E}(z,t)=\frac{w}{c}\sqrt{\frac{8\pi\hbar c^2}{w}}sen(wt-kz)\vec e,
\end{equation}
from where,we deduce that

\begin{equation}
e_0=\frac{w}{c}\sqrt{\frac{8\pi\hbar c^2}{w}}=\sqrt{8\pi\hbar w},
\end{equation}
 where $e_0$ could be thought of as an electric field amplitude
 normalized
 for 1 single photon,with $b_0=e_0$ (Gaussian system),being the
 magnetic
 field amplitude normalized for 1 photon. So we may write

\begin{equation}
\vec{E}(z,t)= e_0sen(wt-kz)\vec e
\end{equation}

 Substituting (8) into (3) and considering the unitary volume
 ($v_{ph}=1$),we obtain

\begin{equation}
\left<E_{eletromag}\right>=\frac{1}{8\pi}e_0^2\equiv\hbar w 
\end{equation} 

  Now,starting from the classical theory of Maxwell for the electromagnetic
 wave,let us consider an average quadratic electric field normalized
 for one
 single photon,which is $e_m=e_0/\sqrt{2}=\sqrt{\left<|\vec
 E|^2\right>}$.
  So doing such a consideration,we may write (9) in the following
 alternative way:

\begin{equation}
\left<E_{eletromag}\right>=\frac{1}{4\pi}e_m^2\equiv\hbar w,
\end{equation}
where we have

\begin{equation}
e_m=\frac{e_0}{\sqrt{2}}=\frac{w}{c}\sqrt{\frac{4\pi\hbar c^2}{w}}
=\sqrt{4\pi\hbar w}
\end{equation}

  It is important to emphasize that,although the
 field in (8)
 is normalized for only one photon,it is still a classical field of
 Maxwell because its value oscillates like a classical wave (solution (8)).
  The only difference is that we have thought about a small amplitude field for
 one photon. Actually the amplitude of the field ($e_0$) cannot be
 measured directly. Only in the classical approximation (macroscopic
 case),
 when we have
 a very large number of photons ($N\rightarrow\infty$),we can somehow
 measure the macroscopic field $E$ of the wave. Therefore,although we
 could
 idealize the case of just one photon as if it were an electromagnetic
 wave of small amplitude,the solution (8) is even a classical one,since
 the field $\vec E$ presents oscillation.

 Actually we already know that the photon wave is a quantum wave,
 i.e.,a de-Broglie wave,where its wavelength ($\lambda=h/p$) is not
 interpreted classically as the oscillation frequency (wavelength due
 to oscillation) of a classical field. However,in a classical case,using the
 solution (8),we would have

\begin{equation}
E_{eletromag}=\frac{1}{4\pi}|\vec{E}(z,t)|^2= 
\frac{1}{4\pi}e_0^2sen^2(wt-kz)
\end{equation}

 In accordance with (12),if the wave of a photon were
 really classical,
 then its energy would not be fixed,as we can see in (12).
  Consequently,its
 energy $\hbar w$ would just be an average value [see (10)]. Hence,in
 order to achieve consistency between the result (10) and the quantum
 wave (de-Broglie wave),we must interpret (10) to be related to the
 de-Broglie wave of the photon with a fixed discrete energy value
 $\hbar w$ instead of an average energy value,since now we consider
 that the
 wave of one single photon
 is a non-classical wave,i.e.,it is a de-Broglie wave. Thus we
 rewrite (10) as follows:

\begin{equation}
E_{eletromag}=E=pc=\frac{hc}{\lambda}=\hbar
 w\equiv\frac{1}{4\pi}e_{ph}^2,
\end{equation} 
where we conclude that

\begin{equation}
\lambda\equiv\frac{4\pi hc}{e_{ph}^2},
\end{equation}
 where $\lambda$ is the
 de-Broglie wavelength. Now,in this case (14),the single
 photon field $e_{ph}$ should not be assumed as a mean value for
 oscillating classical field,and we shall preserve it in order to
 interpret it as
 a {\it scalar quantum electric field \/} (a microscopic field) of a
 photon. So basing on such a heuristic reasoning,let us also call it
 {\it ``scalar support of electric field"},representing a
 quantum (corpuscular)-mechanical aspect of electric field for the
 photon. As $e_{ph}$ is responsible for the energy of the photon
 ($E\propto e_{ph}^2$),where $w\propto e_{ph}^2$ and $\lambda\propto
 1/e_{ph}^2$,indeed we see that $e_{ph}$ presents a quantum behavior,as
 it provides the dual aspects (wave-particle) of the photon,where its
 mechanical momentum may be written as $p=\hbar k=
 2\pi\hbar/{\lambda}$=$\hbar
 e_{ph}^2/2hc$ [refer to (14)],or simply $p=e_{ph}^2/4\pi c$.

\subsection{Electromagnetic nature of the matter}

   Our objective is to extend the idea of the photon electromagnetic
 energy
 [equation (13)] for the matter. By doing this,we shall provide heuristic
 arguments that rely directly on de-Broglie reciprocity postulate,
 which has extended the idea of wave (photon wave) for the matter (electron),
 behaving also like wave. Thus the relation (14) for the photon,which is
 based on de-Broglie relation ($\lambda=h/p$) may also be extended for the
 matter (electron),in accordance with the idea of de-Broglie reciprocity. In
 order to strengthen such an argument,we are going to assume the
 phenomenon of pair formation,where the photon $\gamma$ decays into
 two charged massive particles,namely the electron ($e^{-}$) and its
 anti-particle,the positron ($e^{+}$). Such an example will enable us to better
 understand the need of extending the idea of the photon electromagnetic mass
  ($m_{electromag}=E_{electromag}/c^2$) (equation 13) for the matter ($e^{-}$ and $e^{+}$),
 by using that concept of {\it field scalar support.\/}.

  Now consider the phenomenon of pair formation,i.e.,
$\gamma\rightarrow e^{-}+e^{+}$. Then,by using the conservation of energy
for $\gamma$-decay,we write the following equation:
\begin{equation}
E_{\gamma}=\hbar w = m_{\gamma}c^2=m_0^{-}c^2+m_0^{+}c^2 + K^{-}+
 K^{+}= 
2m_0c^2 + K^{-}+ K^{+},
\end{equation}
where $K^{-}$ and $K^{+}$ represent the kinetic energy for electron and
positron respectively. We have $m_0^{-}c^2=m_0^{+}c^2\cong 0,51$Mev for electron or positron.

Since the photon $\gamma$ electromagnetic energy is
 $E_{\gamma}=h\nu=m_{\gamma}c^2=\frac{1}{4\pi}e_{\gamma}^2$,or else,$E_{\gamma}=\epsilon_0e_{\gamma}^2$
 given in the International
 System of Units (IS),and also knowing that $e_{\gamma}=cb_{\gamma}$
 (IS),where $b_{\gamma}$ represents the {\it magnetic field scalar support}
 for the photon $\gamma$,so we also may write

\begin{equation}
E_{\gamma}=c\epsilon_0(e_{\gamma})(b_{\gamma})
\end{equation}

 Photon has no charge,however,when it is materialized into the pair
 electron-positron,its electromagnetic content given in (16) ceases to
 be free or purely kinetic (purely relativistic mass) to become massive
 through the materialization of the pair. Since such massive particles
 ($v_{(+,-)}<c$) also
 behave like waves in accordance with de-Broglie idea,now it
 would also be natural to extend the relation (14) (of the photon) for
 representing wavelengths of the matter (electron or positron) after the photon-$\gamma$
 decay,namely:

\begin{equation}
\lambda_{(+,-)}\propto\frac{hc}{\epsilon_0[e_s^{(+,-)}]^2}=
\frac{h}{\epsilon_0[e_s^{(+,-)}][b_s^{(+,-)}]},
\end{equation}
where $e_s^{(+,-)}$ and $b_s^{(+,-)}$ play the role of the
electromagnetic content for energy condensed into matter ({\it scalar
 support of electromagnetic field \/} for the matter). Such fields are associated with
the total energy of the moving massive particle,whose mass has
essentially an electromagnetic origin,given in the form

\begin{equation}
m\equiv m_{electromag}\propto e_sb_s,
\end{equation}
where $E=mc^2\equiv m_{electromag}c^2$.

 Basing on (16) and (17),we may write (15) in the following way:

\begin{equation}
E_{\gamma}=c\epsilon_0e_{\gamma}b_{\gamma}=
c\epsilon_0e_s^{-}b_s^{-}v_e^{-}+ c\epsilon_0e_s^{+}b_s^{+}v_e^{+}=
[c\epsilon_0e_{s0}^{-}b_{s0}^{-}v_e + K^{-}]+
[c\epsilon_0e_{s0}^{+}b_{s0}^{+}v_e + K^{+}]=
\end{equation}
~~~~~~~~~~~~~~~~$2c\epsilon_0e_{s0}^{(+,-)}b_{s0}^{(+,-)}v_e+K^{-}+K^{+}=
2m_0c^2 + K^{-} + K^{+}$,\\

where $m_0c^2=m_0^{(+,-)}c^2=c\epsilon_0e_{s0}^{(+,-)}b_{s0}^{(+,-)}v_e\cong
0,51MeV$. $e_{s0}^{(+,-)}$ and $b_{s0}^{(+,-)}$ represent the proper
electromagnetic contents of electron or positron. Later we will
show that the mass $m_0$ does not represent a classic rest mass due
to the inexistence of rest in such a space-time. This question shall be
clarified in 5th section. The volume $v_e$ in (19) is a free variable to be
considered.

  In accordance with equation (19),the present model provides a fundamental
point that indicates electron is not necessarily an exact punctual particle.
 Quantum Electrodynamics,based on Special Relativity deals with the electron as a
punctual particle. The well-known classical theory of the electron foresees for the
electron radius the same order of magnitude of the radius of a proton,i.e.,$R_e\sim 10^{-15}m$.

  The most recent experimental evidence about scattering of electrons
 by electrons at very high kinetic energies indicates that the electron
 can be considered approximately a point particle. Actually electrons
 have an extent less than collision distance,which is about $R_e\sim
 10^{-16}m$\cite{6}. Actually such an extent is negligible in
 comparison to the dimensions of an atom ($10^{-10}m$),or even the dimensions of a
 nucleus ($10^{-14}m$),but it is not exactly a point. By this
 reason,the present model can provide a very small non-null volume $v_e$ for the electron. But,
 if we just consider $v_e=0$ according to (19),we would have an absurd
 result,i.e,divergent internal fields $e_{s0}=b_{s0}\rightarrow\infty$. However,
 for instance,if we consider $R_e\sim 10^{-16}m$ ($v_e\propto R_e^3\sim
 10^{-48}m^3$) for our model,and knowing that $m_0c^2\cong 0,51MeV
 (\sim 10^{-13}J)$,thus,in such a case (see (19)),we would obtain
 $e_{s0}\sim 10^{23}V/m$. Such a value is extremely high and therefore
 we may conclude that
 the electron is extraordinarily compact,with a very high energy
 density. So,for such an example,if we imagine over the ``surface'' of
 the electron,we would detect a field $e_{s0}\sim 10^{23}V/m$ instead of an infinite
 value for it. According to the present model,the field $e_{s0}$ inside the
 almost punctual non-classical electron with such a radius ($\sim 10^{-16}m$)
 would be finite and constant ($\sim10^{23}V/m$) instead of a function like
 $1/r^2$ with divergent classical behavior. Indeed,for $r>R_e$,the field $E$
 decreases like $1/r^2$,i.e,$E=e/r^2$. For $r=R_e$, $E=e/R_e^2\equiv e_{s0}$.
 Actually,for $r\leq R_e$,we have $E\equiv e_{s0}=constant(\sim 10^{23}V/m)$.

 The next section will be dedicated to the investigation about the
 electron coupled to a gravitational field.

\section{Electron coupled to a gravitational field}

 When a photon with energy $h\nu$ is subjected to a certain
 gravitational potential $\phi$,its energy (or frequency) increases to be
 $E^{\prime}=h\nu^{\prime}$,where

\begin{equation}
E^{\prime}=h\nu^{\prime}=h\nu(1+\frac{\phi}{c^2})
\end{equation}

 By convention,as we have stipulated $\phi>0$ to be attractive
potential,we have $\nu^{\prime}>\nu$. By considering (16)
for any photon and by substituting (16) into (20),we alternatively write

\begin{equation}
E^{\prime}=c\epsilon_0 e_{ph}^{\prime}b_{ph}^{\prime}=
c\epsilon_0 e_{ph}b_{ph}\sqrt{g_{00}},
\end{equation}
where $g_{00}$ is the first component of the metric tensor,where
 $\sqrt{g_{00}}=(1+\frac{\phi}{c^2})$ and $e_{ph}=cb_{ph}$.

From (21),we can extract the following relationships,namely:

\begin{equation}
e_{ph}^{\prime}=e_{ph}\sqrt{\sqrt{g_{00}}},~ ~ ~
b_{ph}^{\prime}=b_{ph}\sqrt{\sqrt{g_{00}}}
\end{equation}

  In the presence of gravity,such fields $e_{ph}$ and $b_{ph}$ of the
 photon increase according to (22),leading to the increasing of the
 photon
 frequency or energy,according to (20). Thus we may think about the
 following
 increments,namely:

\begin{equation}
\Delta e_{ph}=e_{ph}^{\prime}-e_{ph}=e_{ph}(\sqrt{\sqrt{g_{00}}} - 1),~
 ~
\Delta b_{ph}=b_{ph}^{\prime}-b_{ph}=b_{ph}(\sqrt{\sqrt{g_{00}}} - 1)
\end{equation}

  In accordance with General Relativity (GR),when a massive particle
 of mass $m_0$ moves in the presence of a gravitational potential $\phi$,its total
 energy $E$ is given in the following way:

\begin{equation}
E=mc^2=m_0c^2\sqrt{g_{00}} + K,
\end{equation}
where we can think that $m_0(=m_0^{(+,-)})$ is the mass of the
electron or positron,emerging from $\gamma$-decay in the presence
of a gravitational potential $\phi$.

 In order to facilitate the understanding of what we are proposing,let
 us consider $K<<m_0c^2$,since we are interested only in obtaining the influence
 of the potential $\phi$. Therefore we write

\begin{equation}
E=m_0c^2\sqrt{g_{00}}
\end{equation}

 As we already know that $E_0=m_0c^2=c\epsilon_0e_{s0}^{(+,-)}b_{s0}^{(+,-)}v_e$,we can
 also write the total energy $E$,as follows:

\begin{equation} 
E =c\epsilon_0 e_s^{(+,-)}b_s^{(+,-)}v_e=
c\epsilon_0 e_{s0}^{(+,-)}b_{s0}^{(+,-)}v_e\sqrt{g_{00}},
\end{equation}
from where we can extract

\begin{equation}
e_s^{(+,-)}=e_{s0}^{(+,-)}\sqrt{\sqrt{g_{00}}},~ ~ ~
b_s^{(+,-)}=b_{s0}^{(+,-)}\sqrt{\sqrt{g_{00}}}.
\end{equation}

 So we obtain

\begin{equation}
\Delta e_s=e_{s0}^{(+,-)}(\sqrt{\sqrt{g_{00}}} - 1),~ ~
\Delta b_s=b_{s0}^{(+,-)}(\sqrt{\sqrt{g_{00}}} - 1), 
\end{equation}
where we have $\Delta e_s=c\Delta b_s$.

 As the energy of the particle can be represented as a condensation of
electromagnetic fields in scalar forms $e_s$ and $b_s$,this model is
 capable of assisting us to think that the well-known external fields
 $\vec{E}$ and
 $\vec{B}$ for the moving charged particle,by storing an energy density
($\propto |\vec{E}|^2 +|\vec{B}|^2$) should also suffer some influence
 (shifts) in the presence of
 gravitational potential. In accordance with GR,every kind of energy
 is also a source of gravitational field. This non-linearity that is
 inherent to the gravitational field leads us to think that,at least in a certain
 approximation in the presence of gravity,the external fields $E$ and
 $B$ should experiment positive small shifts $\delta E$ and $\delta B$,which are
 proportional to the intrinsic increments (shifts) $\Delta e_s$ and $\Delta b_s$ of
 the particle,namely:

\begin{equation}
\delta E=(E^{\prime}-E)\propto\Delta e_s=(e_s-e_{s0}),~ ~
\delta B=(B^{\prime}-B)\propto\Delta b_s=(b_s-b_{s0})
\end{equation}

Here we have omitted the signs $(+,-)$ in order to simplify the notation.
 Since $\Delta e_{s}=c\Delta b_{s}$,then $\delta E=c\delta B$.

  In accordance with (29),we may conclude that there is a constant of
 proportionality that couples the external electromagnetic fields $E$
 and $B$ of the moving charge with gravity by means of the small
 shifts $\delta E$ and $\delta B$. Such a constant works like a fine-tuning,namely:

\begin{equation}
\delta E=\xi\Delta e_s,~ ~
\delta B=\xi\Delta b_s,
\end{equation}
where $\xi$ is a dimensionaless constant to be obtained. We expect that
$\xi<<1$ due to the fact that the gravitational interaction is much
weaker than the electromagnetic one. $\delta E$ and $\delta B$
depend only on $\phi$ over the electron.

 Substituting (28) into (30),we obtain

\begin{equation}
\delta E=\xi e_{s0}(\sqrt{\sqrt{g_{00}}} - 1), ~ ~ 
\delta B=\xi b_{s0}(\sqrt{\sqrt{g_{00}}} - 1).   
\end{equation}

 Due to the very small positive shifts $\delta E$ and $\delta B$ in the
 presence of a weak gravitational potential $\phi$,the total electromagnetic
 energy density in the space around the charged particle is slightly
 increased,as follows:

\begin{equation}
\rho_{electromag}^{total}=\frac{1}{2}\epsilon_0(E+\delta
 E)^2+\frac{1}{2\mu_0}
(B +\delta B)^2
\end{equation}

Substituting (31) into (32) and performing the calculations,we will finally obtain

\begin{equation}
\rho_{electromag}^{total}=\frac{1}{2}[\epsilon_0
 E^2+\frac{1}{\mu_0}B^2] +
\xi[\epsilon_0 Ee_{s0}+\frac{1}{\mu_0}Bb_{s0}](\sqrt{\sqrt{g_{00}}} -
 1)+\frac{1}{2}\xi^2[\epsilon_0 (e_{s0})^2+\frac{1}{\mu_0}(b_{s0})^2](\sqrt{\sqrt{g_{00}}}-1)^2
\end{equation}

We may assume that $\rho_{electromag}^{total}=\rho_{electromag}^{(0)}+
\rho_{electromag}^{(1)}+\rho_{electromag}^{(2)}$ for representing (33),
 where the first term $\rho_{electromag}^{(0)}$ is the free electromagnetic energy density
 (zero order) for the ideal case of a charged particle uncoupled from gravity
 ($\xi=0$),i.e,the ideal case of a free particle (a perfect plane
 wave,which does not exist in reality due always to the presence of gravity).
  We have $\rho^{(0)}\propto 1/r^4$ ({\it coulombian term \/}).

The coupling term $\rho^{(1)}$ (second term) represents an electromagnetic energy
density of first order,that is,it contains an influence of 1st order for $\delta E$ and
$\delta B$,as it is proportional to $\delta E$ and $\delta B$ due
to a certain influence of gravity. Therefore it is a mixture term that behaves
essentially like a {\it radiation term}. Thus we have $\rho^{(1)}\propto
1/r^2$,since $e_{s0}$ (or $b_{s0}$) $\sim constant$ and $E$(or $B$)$\propto 1/r^2$.
 It is very interesting to notice that such a radiation term of a charge in a true gravitational
field corresponds effectively to a certain radiation field due to an slightly accelerated charge
in free space,however such an equivalence is weak due to the very small value of $\xi$.

 The last coupling term ($\rho^{(2)}$) is purely interactive due to the presence of gravity only.
 This means that it is a 2nd order interactive electromagnetic energy density term,since it is proportional to
 $(\delta E)^2$ and to $(\delta B)^2$. Hence we have $\rho^{(2)}\propto 1/r^0\sim constant $,being
 $\rho^{(2)}=\frac{1}{2}\epsilon_0(\delta E)^2+\frac{1}{2\mu_0}(\delta B)^2=\epsilon_0(\delta E)^2=
 \frac{1}{\mu_0}(\delta B)^2$,which varies only with the gravitational potential
($\phi$). Since we have $\rho^{(2)}\propto 1/r^0$,it has a non-locality behavior. This means that $\rho^{(2)}$
 behaves like a kind of non-local field,that is inherent to the
 space ({\it a constant term for representing a background field\/}). It does not depend on the distance $r$
from the charged particle. So it is a constant energy density for a fixed potential $\phi$,and fills the whole space.
 $\rho^{(2)}$ always exists due to the inevitable presence of gravity and therefore it cannot be cancelled. Due to this fact,the increment
$\delta B$ that contributes for the density of interactive energy $\rho^{(2)}$ cannot
vanish since the electron is not free ($\rho^{(2)}\neq 0$). This always assures a non-zero value of magnetic field ($\delta B\neq 0$)
for any transformation,and so this is the fundamental reason why the fields $E$ and $B$ should coexist in the
presence of gravity,where the charge experiments a background field ($\rho^{(2)}\propto(\delta B)^2$)
 connected to a privileged reference
frame of an unattainable minimum speed that justifies in a kinematic point of view the impossibility of finding
$\delta B=0$. This minimum speed ($V$) is a universal constant that should be related directly to gravity ($G$),
since $V$ is also responsible for the coexistence of $E$ and $B$. We will see such a connection in the next section.

   Usually we have $\rho^{(0)}>>\rho^{(1)}>>\rho^{(2)}$.
  For a very weak gravitational field,we can consider a good practical
 approximation as $\rho_{eletromag}^{total}\approx\rho^{(0)}$. However,from a
 fundamental point of view,we cannot neglect the coupling terms,specially
 the last one for large distances,as it has a vital importance in this
 work,permiting us to understand a non-local vacuum energy that is
 inherent to the space,i.e. $\rho^{(2)}\propto 1/r^0$. Such a
 background field with
energy density $\rho^{(2)}$ has deep implications for our
 understanding of the space-time structure at very large scales of
 length (cosmological scales),since $\rho^{(2)}$ does not have
 $r$-dependence,i.e,it remains for $r\rightarrow\infty$.

  In the next section,we will estimate the constant $\xi$ and
 consequently the idea of a universal minimum velocity in the space-time. Its
cosmological implications will be treated in section 8.

\section{The fine adjustment constant $\xi$ and its implications}

   Let us begin this section by considering the well-known problem that
 deals
 with the electron at the bound state of a coulombian potential of a
 proton
 (Hydrogen atom). We start from this subject because it poses a
 certain
 similarity with the present model for the electron coupled to a
 gravitational
 field. We know that the fine structure constant ($\alpha_F=1/137$)
 plays an
 important role for obtaining the energy levels that bond the electron to
 the
 nucleus (proton) in the Hydrogen atom. Therefore,in a similar way
 to such an
 idea,we plan to extend it in order to see that the fine coupling
 constant
$\xi$ plays an even more fundamental role than the fine structure
 $\alpha_F$,by
 considering that $\xi$ couples gravity to the electromagnetic field of
 the electron charge.

Let's initially consider the energy that bonds the electron to the
proton at the fundamental state of the Hydrogen atom,as follows:

\begin{equation}
\Delta E=\frac{1}{2}\alpha_F^2m_0c^2,
\end{equation}
where $\Delta E$ is assumed as module. We have $\Delta E<<m_0c^2$,
 where $m_0$
 is the electron mass,which is practically the reduced mass of the
 system ($\mu\approx m_0$).

 We have $\alpha_F=e^2/\hbar c=q_e^2/4\pi\epsilon_0\hbar c\approx1/137$
 (fine structure constant). Since $m_0c^2\cong 0.51$ Mev,we have
$\Delta E\approx 13.6$eV.

Since we already know that $E_0=m_0c^2=c\epsilon_0e_{s0}b_{s0}v_e$,so
we may write (34) in the following alternative way:

\begin{equation}
\Delta E=\frac{1}{2}\alpha_F^2c\epsilon_0e_{s0}b_{s0}v_e= 
\frac{1}{2}c\epsilon_0(\alpha_F e_{s0})(\alpha_F b_{s0})v_e\equiv
\frac{1}{2}c\epsilon_0(\Delta e_{s})(\Delta b_{s})v_e,
\end{equation}
from where we extract

\begin{equation}
\Delta e_{s}\equiv\alpha_F e_{s0},~~
\Delta b_{s}\equiv\alpha_F b_{s0}.
\end{equation}

 It is interesting to observe that (36) maintains a certain
 similarity with (30),however,first of all,we must emphasize that
 the variations $\Delta e_s$ and $\Delta b_s$ for the electron energy
 have a purely coulombian origin,since the fine structure constant $\alpha_F$
 depends solely on the electron charge. Thus we can write the electric force
 between two electronic charges in the following way:

\begin{equation}
F_e=\frac{e^2}{r^2}=\frac{q_e^2}{4\pi\epsilon_0 r^2}=\frac{\alpha_F
\hbar c}{r^2},
 \end{equation}
where $e=q_e/\sqrt{4\pi\epsilon_0}$.

 If we just consider a gravitational interaction between two electrons,we would have

\begin{equation}
F_g=\frac{Gm_e^2}{r^2}=\frac{\beta_F\hbar c}{r^2},
 \end{equation}
from where we obtain

\begin{equation}
\beta_F=\frac{Gm_e^2}{\hbar c}.
 \end{equation}

 We have $\beta_F<<\alpha_F$ due to the fact that the gravitational interaction
 is much weaker than the electric one,so that
 $F_e/F_g=\alpha_F/\beta_F\sim 10^{42}$,where $\beta_F\cong 1.75\times
 10^{-45}$. Therefore we shall call $\beta_F$\footnote{we must not
 mistake
 superfine structure $\beta_F$ with hyperfine structure (spin
 interaction), as
 they are completely different.} the {\it superfine structure constant
 \/},since gravitational interaction creates a bonding energy
 extremely smaller
 than the coulombian bonding energy considered for the fundamental
 state ($\Delta E$) in the Hydrogen atom.

 To sum up,whereas $\alpha_F(e^2)$ provides the adjustment for the
 coulombian bonding energies between two electronic charges,
 $\beta_F(m_e^2)$ gives the adjustment for the gravitational
 bonding
 energies between two electronic masses.
  Such bonding
 energies of electrical or gravitational origin increment the particle
 energy
 through $\Delta e_s$ and $\Delta b_s$.

 Now,following the above reasoning,we notice that the present model
 enables us to
 introduce the very fine-tuning (coupling) $\xi$ between gravity (a
 gravitational potential generated by the mass $m_e$)
 and electrical field (electrical energy density generated by the
 charge $q_e$ (refer to (30))). Thus for such more fundamental
 case,we have a kind of bond of the type $m_eq_e$ (mass-charge) through the
 adjustment (coupling) $\xi$. So the subtleness here is that the bonding energy
 density due to $\xi$,by means of the increments $\delta E$ and $\delta B$ (see (30),
 (31),(32) or (33)) occurs on the electric and magnetic fields generated
 in the space by the own charge $q_e$.

Although we could show a laborious and step by step problem for
 obtaining the
 constant $\xi$,the way we follow here is shorter because it starts
 from
 important analogies by using the ideas of fine structure $\alpha_F=
 \alpha_F(e^2)$,i.e.,an eletric interaction ({\it charge-charge \/})
 and also superfine structure $\beta_F=\beta_F(m_e^2)$,i.e.,a gravitational
 interaction ({\it mass-mass \/}). Hence,now it is easy to conclude
 that the
 kind of mixing coupling we are proposing,of the type ``$m_eq_e$''
 ({\it
 mass-charge\/}) represents a gravi-electrical coupling constant,which
 leads
 us naturally to think that such a constant $\xi$ is of the form
 $\xi=\xi(m_eq_e)$,and therefore meaning that

\begin{equation}
\xi=\sqrt{\alpha_F\beta_F},
\end{equation}
which represents a geometrical average between electrical and
 gravitational
couplings,and so we finally obtain from (40)

\begin{equation}
\xi=\sqrt{\frac{G}{4\pi\epsilon_0}}\frac{m_eq_e}{\hbar c},
\end{equation}
where indeed we have $\xi=\xi(m_eq_e)\propto m_eq_e$. From (41) we obtain
 $\xi\cong 3.57\times 10^{-24}$. Let us call $\xi$ {\it fine adjustment
 constant.} The quantity $\sqrt{G}m_e$ in (41) can be thought of as a
 {\it gravitational charge} $e_g$,so that $\xi=e_ge/\hbar c$.

In the Hydrogen atom,we have the fine structure constant
 $\alpha_F=e^2/\hbar c=
v_B/c$, where $v_B=e^2/\hbar=c/137$. This is the velocity of the
 electron at the
 atom fundamental level (Bohr velocity). At this level,the electron
 does not
 radiate because it is in a kind of balance state,in spite of its
 electrostatic interaction with the nucleus (centripete force),namely
 it works effectively like an inertial system. Hence,
 following an
 analogous reasoning for the more fundamental case of the constant
 $\xi$,we may also write (41) as the ratio of two velocities,as follows:

\begin{equation}
\xi=\frac{V}{c},
\end{equation}
from where we have

\begin{equation}
V=\xi c=\frac{e_ge}{\hbar}=\sqrt{\frac{G}{4\pi\epsilon_0}}\frac{m_eq_e}{\hbar},
\end{equation}
where $V\cong 1.07\times 10^{-15}m/s$. In the newtonian (classical)
 universe,where $c\rightarrow\infty$ and $V\rightarrow 0$,we have
 $\xi\rightarrow 0$. So the coupling of fields is impossible. Under
Einstein's theory (relativistic theory),$V\rightarrow 0$ and we also
have $\xi\rightarrow 0$,where,although electrodymanics is compatible with
 relativistic mechanics,gravitation is still not properly coupled to
 electrodynamics at quantum level. In the present model that breaks
 Lorentz
 symmetry,where $\xi\sim 10^{-24}$,gravitation is coupled to
 electrodynamics of moving particles. The quantum uncertainties
 should naturally arise from such a symmetric space-time structure ($V<v\leq
 c$),which will be denominated {\it Symmetrical Special Relativity} (SSR)
due to the existence of two limits of speed.

Similarly to the Bohr velocity ($v_B$) for fundamental bound state,the
 speed $V$ is also a universal fundamental constant,however the crucial
 difference
 between them is that $V$ is associated with a more fundamental bound
 state in the Universe as a whole,since
 gravity ($G$),which is the weakest interaction plays now an important
 role for
 the dynamics of the electron (electrodynamics) in such a space-time.
 This may be observed in (43) because,if we make $G\rightarrow 0$,we would
 have $V\rightarrow 0$ and so we will recover the case of the classical vacuum
 (empty space or no background field).

Our aim is to postulate $V$ as an unattainable universal (constant)
minimum speed associated with a privileged frame of background field,but before
 this,we
 must provide a better
 justification of why we consider the electron mass and charge to
 calculate
 $V$ $(V\propto m_eq_e)$,instead of masses and charges of other
 particles.
  Although there are fractionary electric charges as the case
 of quarks,such charges are not free in Nature for bonding only with
 gravity. They
 are strongly connected by the strong force (gluons). Actually the
 charge of
 the electron is the smallest free charge in Nature. Besides this,
 the electron is the elementary charged particle with the smallest
 mass. Therefore the product $m_eq_e$ assumes a minimum value. And in
 addition to that,
 the electron is completely stable. Other charged particles such as for
 instance
 $\pi^{+}$ and $\pi^{-}$ have masses that are greater than the electron mass,
 and they are unstable,decaying very quickly. Such a subject may
 be dealt with more extensively in another article.

 We could think about a velocity $Gm_e^2/\hbar$ $(<<V)$ that has origin
 from a purely
gravitational interaction,however such a much lower bound state does
 not exist because
the presence of electromagnetic interactions is essential at subatomic
 level. And since
neutrino does not interact with electromagnetic field,it cannot be
 considered
to estimate $V$.

Now we can verify that the minimum speed ($V$) given in (43) is
directly related to the minimum length of quantum gravity (Planck
length),as follows:

\begin{equation}
V=\frac{\sqrt{G}m_e e}{\hbar}=(m_ee\sqrt{\frac{c^3}{\hbar^3}})l_p, 
\end{equation}
where $l_p=\sqrt{G\hbar/c^3}$.

 In (44),as $l_p$ is directly related to $V$,
 if we make $l_p\rightarrow 0$ by considering $G\rightarrow 0$,this
 implies $V\rightarrow 0$ and thus we restore the case of the classical space-time in Relativity.

   Now we can notice that the universal constant of minimum speed $V$
 in (44),associated with very low energies
(very large wavelengths) is directly related to the universal constant
 of minimum length
 $l_p$ (very high energies),whose invariance has been studied in DSR
 by Magueijo,Smolin,Camelia et al [20-25].

 The natural consequence of the presence of a more fundamental level
 associated with $V$ in the space-time is the existence
 of a privileged reference frame of
 background field in the Universe. Such a frame
 should be connected to a kind of vacuum energy,that is inherent to the space-time
(refer to $\rho^{(2)}$ in equation (33)). This
 idea reminds us of the conceptions of Mach\cite{2},
 Schroedinger\cite{3} and
 Assis\cite{7},although such conceptions are still within the
 classical context.

  Since we are assuming an absolute and privileged reference frame
 ($V$),
 which is
 underlying and also inherent to the whole space-time geometry,we shall
 call
 it ultra-referential-$S_V$. By drawing inspiration from some of the
 non-conventional ideas of Einstein in relation to the
 ``ether''\cite{8},let
 us assume that such an ultra-referential of background field
 $S_V$,which in a way redeems
 his ideas,introduces a kind of relativistic ``ether" of the
 space-time. Such a
 new concept has nothing to do with the so-called luminiferous ether
(classical ether) established before Relativity theory.

    The present idea about a relativistic ``ether'' for the
 ultra-referential
 $S_V$ aims at the implementation of the quantum principles
 (uncertainties) in
 the space-time. This line of investigation
 resumes those non-conventional Einstein's ideas
 \cite{8}\cite{9},who attempted to bring back the idea of a new
 ``ether" that cannot be
 conceived as composed of punctual particles and having a world
 line followed in the time.

 Actually such an idea of ``ether'' as conceived by Einstein should be
 understood
 as a non-classical concept of ether due essentially to its
 non-locality
 feature. In this sense, such a new ``ether'' has a certain
 correspondence with
 the ultra-referential $S_V$ due to its totality as a physical space,
 not
 showing any movement. In fact,as $S_V$ would be absolutely
 unattainable
 for all
 particles (at local level),$V$ would prohibits to think about a
perfect plane wave ($\Delta x=\infty$),since it is an idealized case
 associated
with the perfect equilibrium of a free particle ($\Delta p=0$). So the
ultra-referential $S_V$ would
 really be non-local ($\Delta x=\infty$),which is in agreement with
 that Einstein's
 conception about an ``ether'' that could not be split into isolated
 parts and
 which,due
 to its totality in the space,would give us the impression that it is
 actually
stationary. {\it In order to understand better its non-locality feature
 by
using a symmetry reasoning,we must perceive that such a minimum limit
 $V$ works
in a reciprocal way when compared with the maximum limit $c$,so that
 particles supposed in such a limit $V$,in contrast of what would
 happen in
 the limit $c$,would become completely `` defrosted '' in the space
 ($\Delta x
\rightarrow\infty$) and time ($\Delta\tau\rightarrow\infty$),being in
 anywhere
in the space-time and therefore having a non-local behavior. This super ideal
condition corresponds to the ultra-referential $S_V$},at which
the particle would have an infinite de-Broglie wavelength,being completely
spread out in the whole space. This state coincides with the background field for $S_V$,
however $S_V$ is unattainable for all the particles.

  In vain,Einstein attempted to satisfactorily redeem the idea of a
 new ``ether"
 under Relativity in various manners\cite{9,10,11,12,13,14}
 because,in effect,his theory wasn't still able to adequately implement the
 quantum uncertainties as he also tried to do\cite{15,16,17},and in
 this respect,Relativity is still a classical theory,although the new
 conception of ``ether" presented a few non-classical characteristics. Actually it was
 Einstein who coined the term {\it ultra-referential \/} as the fundamental
 aspect of Reality. To him,the existence of an ultra-referential
 cannot be
 identified with none of the reference frames in view of the fact that
 it is a
 privileged one in respect of the others. This seems to
 contradict
 the principle of Relativity,but,in vain,Einstein attempted to find
 a relativistic ``ether" (physical-space),that is inherent to the geometry
 of the
 space-time,which does not contradict such a principle. That was the
 problem
 because such a new ``ether" does not behave like a Galilean reference
 frame and,
 consequently,it has nothing to do with that absolute space filled
 by the
 luminiferous ether,although it behaves like a privileged background
 field in the Universe.

 The present work seeks to naturally implement the quantum principles
 into the space-time. Thanks to the current investigation,we shall notice that
 Einstein's non-conventional ideas about the relativistic ``ether" and also his
 vision\cite{18} of making quantum principles to emerge naturally from a
 unified field theory become closely related between themselves.

\section{A new conception of reference frames and space-time
 interval: a fundamental explanation for the uncertainty principle}

\subsection{Reference frames and space-time interval}

  The conception of background privileged reference frame
 (ultra-referential
 $S_V$) has deep new implications for our understanding of reference
 systems. That classical notion we have about the inertial (Galilean)
 reference
 frames,where the idea of rest exists, is eliminated at quantum level,
 where gravity plays a fundamental role for such a space-time with a
 vacuum energy associated with $S_V$ ($V\propto G^{1/2}/\hbar$).

  Before we deal with the implications due to the implementation
 of such a
 ultra-referential
 $S_V$ in the space-time at quantum level,let us make a brief
 presentation of the
 meaning of the
 Galilean reference frame (reference space),well-known in Special
 Relativity. In accordance with that theory,when an observer assumes
 an infinite number of points at rest in relation to himself,he
 introduces his own reference space $S$. Thus,for another observer
 $S^{\prime}$
who is moving with a speed $v$ in relation to $S$, there should also
 exist an
 infinite number of points at rest at his own reference frame.
 Therefore,
 for the observer $S^{\prime}$,the reference space $S$ is not standing
 still and it has its points moving at a speed $-v$. For this reason,
in accordance with the principle of relativity,there is no
privileged Galilean reference frame at absolute rest,since the reference
space of a given observer becomes movement for another one.

The absolute space of pre-einsteinian physics,connected to
 the ether
 in the old sense,also constitutes by itself a reference space. Such a
space was assumed as the privileged reference space of the absolute
 rest. However,as it was also essentially a Galilean reference space
 like
 any other,
 comprised of a set of points at rest,actually it was also subjected to
 the notion of movement. The idea of movement could be applied to the
 ``absolute space'' when,for instance,we assume an observer on Earth,
 which is moving
 with a speed $v$ in relation to such a space. In this case,for an
 observer at
 rest on Earth,the points that would constitute the absolute space of
 reference would be moving at a speed of $-v$. Since such an absolute
 space was
 connected to the old ether,the Earth-bound observer should detect a
 flow
 of ether
$-v$,however the Michelson-Morley experiment has not detected such an
 ether.

Einstein has denied the existence of the ether associated with a
 privileged
 reference frame because it has contradicted the principle of
 relativity.
 Therefore
 this idea of a Galilean ether is superfluous,as it would also merely
 be a reference space constituted by points at rest,as well as any
 other.
  In this respect,there is nothing special in such a classical
 (luminiferous) ether.

However,motivated by the provocation from H. Lorentz and Ph. Lenard
 Lorentz\cite{8},Einstein attempted to introduce several new
 conceptions of
 a new ``ether'',which did not contradict the principle of relativity.
 After
 1925,he started using the word ``ether'' less and less
 frequently,although he
 still wrote in 1938:{\it ``This word `ether' has changed its meaning
 many
 times,in the development of Science... Its history,by no means
 finished,is
 continued by Relativity theory\cite{10}... ''\/}.

In 1916,after the final formulation of GR,Einstein proposed a
 completely new concept of ether. Such a new ``ether" was a
 relativistic
 ``ether",which described space-time as a {\it sui generis \/} material
 medium,which in no way could constitute a reference space subjected to the
 relative notion
 of movement. Basically,the essential characteristics of the new
 ``ether" as
 interpreted by Einstein can be summarized as follow:

-{\it It constitutes a fundamental ultra-referential of
 Reality,which is
 identified with the physical space,being a relativistic ether,i.e.,
 it is
covariant because the notion of movement cannot be applied to it,which
 represents a kind of absolute background field that is inherent to the
 metric $g_{\mu\nu}$ of the space-time\/}.

-{\it It is not composed of points or particles,therefore it cannot be
 understood as a Galilean reference space for the hypothetical
 absolute space.
  For this reason,it does not contradict the well-known principle of
 Relativity\/}.

-{\it It is not composed of parts,thus its indivisibility reminds the
 idea of non-locality\/}.

-{\it It constitutes a medium which is really incomparable with any
 ponderable
 medium constituted of particles,atoms or molecules. Not even the
 background cosmic
 radiation of the Universe can represent exactly such a medium as an
 absolute
 reference system (ultra-referential)\cite{19} \/}.

-{\it It plays an active role on the physical
 phenomena\cite{11}~\cite{12}.} In
 accordance with Einstein,it is impossible to formulate a complete
 physical
 theory without the assumption of an ``ether''(a kind of non-local
 vacuum
 field),
 because a complete physical theory must take into
 consideration
 real properties of the space-time.

 The present work attempts to follow this line of reasoning that
 Einstein did
 not finish,providing a new model with respect to the fundamental
 idea of unification,namely the electrodynamics of a charged particle
  (electron) moving in a gravitational field.

 As we have interpreted the lowest limit $V$ (formulas (43) and (44))
 as unattainable and constant (invariant),such a limit should be
 associated with a
 privileged non-Galilean reference system,since $V$ must remain
 invariant for
 any frame with $v>V$. As a consequence of such a covariance of the
 relativistic ``ether" $S_V$,new speed transformations will show
 that it is
 impossible to cancel the speed of a particle over its own reference
 frame,in
 such a way to always preserve the existence of a magnetic field
 $\vec B$
 for such a charged particle. Thus we should have a speed
 transformation that will show us that $``v-v''>V$ for $v>V$ (see section 6),
where the constancy of $c$ remains,i.e.,$``c-c''=c$ for $v=c$.

  Since it is impossible to find with certainty the rest for a given
 non-Galilean reference system $S^{\prime}$ with a speed $v$ with respect to the
 ultra-referential $S_V$,i.e., $``v-v''\neq 0(>V)$ (section 6),consequently it is also
impossible to find by symmetry a speed $-v$ for the
relativistic ``ether'' when an ``observer'' finds himself at the reference
system $S^{\prime}$ assumed with $v$. Hence,due to such an asymmetry,the flow $-v$ of
the ``ether'' $S_V$ does not exist and therefore,in this sense,it mantains covariant
 ($V$). This asymmetry breaks that equivalence by exchange of reference
 frame $S$ for $S^{\prime}$ through an inverse transformation. Such a
 breakdown of symmetry by an inverse transformation breaks Lorentz
 symmetry due to the presence of the background field for $S_V$ (section 6).

 There is no Galilean reference system in such a space-time,
 where the ultra-referential $S_V$ is a non-Galilean reference system and in
 addition a privileged one (covariant),exactly as is the speed of light
 $c$. Thus the new transformations of speed shall also show that
 $``v\pm V''=v$ (section 6) and $``V\pm V''=V$ (section 6).

 Actually,if we make $V\rightarrow 0$,we therefore recover the
 validity of the
 Galilean reference frame of Special Relativity (SR),where only the
 invariance
 of $c$ remains. In this classical case (SR),we have reference systems
 constituted by a set of points at rest or essentially by macroscopic
 objects. Now,it is interesting to notice that SR contains
 two postulates which conceptually exclude each other in a certain
 sense,namely:

 1) -{\it the equivalence of the inertial reference frames (with $v<c$)
 is essentially
 due to the fact that we have Galilean reference frames,where $v_{rel}=v-v=0$,
 since it is always possible to introduce a set of points at relative
 rest and,consequently,for this reason,we can exchange $v$ for $-v$ by symmetry
 through inverse transformations.}

 2) -{\it the constancy of $c$,which is unattainable by massive
 particles
 and therefore it could never be related to a set of infinite points
 at relative rest. In this sense, such ``referential''($c$),contrary to
 the 1st. one,is not Galilean because we have $``c-c''\neq 0$ $(=c)$
 and,for this reason,we can never exchange $c$ for $-c$.}

 However,the covariance of a relativistic ``ether" $S_V$ places the
 photon ($c$) in a certain condition of equality with the motion of
 other
 particles ($v<c$),just
 in the sense that we have completely eliminated the classical idea of
 rest for
 reference space (Galilean reference frame) in such a space-time.
  Since we
 cannot think about a reference system constituted by a set of infinite
 points at rest in such a space-time,we should define a
 non-Galilean
 reference system essentially as a set of all those particles which
 have the
 same state of motion ($v$) in relation to the ultra-referential-$S_V$ of the
 relativistic ``ether". Thus SSR should contain 3 postulates as follow:

 1) -{\it the constancy of the speed of light ($c$)}.

 2) -{\it the non-equivalence (asymmetry) of the non-Galilean reference
 frames,
 i.e.,we cannot exchange $v$ for $-v$ by the inverse transformations,
 since $``v-v''>V(\propto\sqrt{G}/\hbar)$}, which breaks Lorentz
 symmetry
due to the universal background field associated with $S_V$.

 3) -{\it the covariance of a relativistic ``ether" (ultra-referential $S_V$) associated with
 the unattainable minimum limit of speed $V$.}

 The three postulates described above are compatible among themselves,
 in the sense that we completely eliminate any kind of Galilean
 reference system for the space-time of SSR.

  Figure 1 illustrates a new conception of reference systems in SSR.
\begin{figure}
\includegraphics[scale=1]{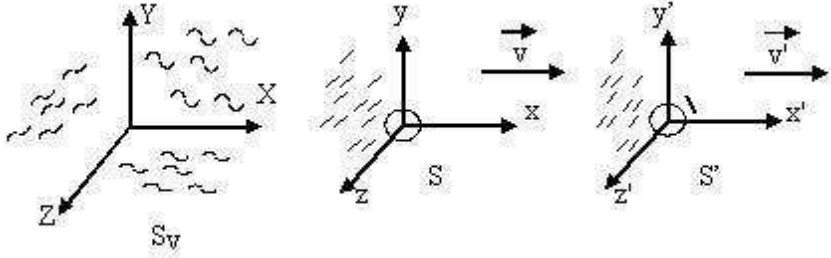}
\caption{$S_V$ is the covariant ultra-referential of background field
 (relativistic ``ether"). $S$ represents the non-Galilean reference frame
 for a massive particle with
 speed $v$ in relation to $S_V$,where $V<v<c$.
 $S^{\prime}$ represents the non-Galilean reference frame for a massive
 particle with speed $v^{\prime}$ in relation to $S_V$. In this
 instance,we
 consider $V<v\leq v^{\prime}\leq c$.\/}
\end{figure}

 Under SR,there is no
 ultra-referential $S_V$,i.e.,$V\rightarrow 0$. Hence,the starting
 point for
 observing $S^{\prime}$ is the reference frame $S$,at which the
 classic observer thinks he is at rest (Galilean reference frame $S$).

 Under SSR,the starting point for obtaining
 the actual motion of all particles of $S^{\prime}$ is the
 ultra-referential $S_V$ (see Fig.1). However,due to the non-locality of $S_V$,that is
 unattainable by the particles,the existence of an observer
 (local level) at it ($S_V$) becomes inconceivable. Hence,let us think
 about a non-Galilean frame $S_0$ for a certain intermediate speed mode
 ($V<<v_0<<c$) in order to represent the starting point at local
 level for ``observing" the motion of $S^{\prime}$ across
 the ultra-referential $S_V$. Such a frame $S_0$
 (for $v_0$ with respect to $S_V$) plays the similar role of a
 ``rest'',in the sense that we restore all the
 newtonian parameters of the particles,such as the proper time interval
 $\Delta\tau$,i.e.,$\Delta t$($v=v_0$)=$\Delta\tau$,the mass $m_0$,
 i.e.,$m(v=v_0)= m_0$,among others. Therefore $S_0$ plays a
 role that is similar to the frame $S$ under SR,where $\Delta
 t(v=0)=\Delta\tau$,$m(v=0)=m_0$, etc. However,here in SSR,the
 classical relative
 rest ($v=0$) of $S$ should be replaced by a universal ``quantum rest'' $v_0(\neq
 0)$ of the non-Galilean frame $S_0$. We will show that
 $v_0$ is also a universal constant.
  {\it In short,$S_0$ is a universal non-Galilean reference frame with speed
 $v_0$ given with respect to $S_V$. At $S_0$,the well-known proper mass ($m_0$)
or proper energy $E_0=m_0c^2$
of a particle is restored}. This means that,at such a frame $S_0$,
we have the proper energy $E=E_0=m_0c^2=m_0c^2\Psi(v_0)$,such that
$\Psi(v_0)=1$,as well as
 $\gamma(v=0)=1$ for the particular case of Lorentz transformations,where
$V\rightarrow 0$. So we will look for the general function $\Psi(v)$ of
 SSR,where we have $E=m_0c^2\Psi(v)$. In the limit $V\rightarrow 0$,
indeed we expect that the function $\Psi(v)\rightarrow\gamma(v)=(1-v^2/c^2)^{-1/2}$
 (see Fig.7).

 By making the non-Galilean reference frame $S$ (Fig.1) coincide with
 $S_0$,we get Figure 2.

\begin{figure}
\includegraphics[scale=0.85]{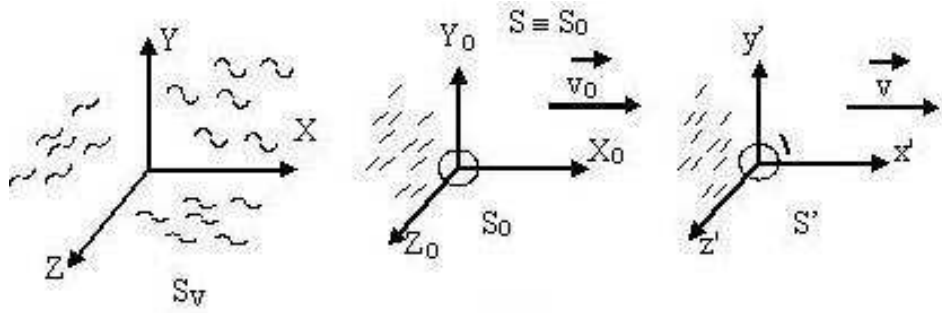}
\caption{\it As $S_0$ is fixed (universal),being $v_0 (>>V)$ given with
 respect
 to $S_V$,we should also consider the new
 interval $V~(S_V)<v~(S^{\prime})\leq v_0~(S_0)$. This non-classical
 regime for $v$
 introduces
 a new symmetry in the space-time,leading to SSR. Thus we expect that
 new and
 interesting results take place. In such an interval ($V<v\leq v_0$),we will
 see that $0<\Psi(v)\leq 1$ (see equations (60),(72) and Fig.7).}
\end{figure}

In general,we should have the total interval $V<v<c$ for $S^{\prime}$ (Fig.2). In
 short we say
 that both of the frames $S_V$ and $S_0$ are already fixed or
 universal,whereas
 $S^{\prime}$ is
 a rolling frame to describe the variations of the moving state
 of the particle within such a total interval. Since the rolling frame
 $S^{\prime}$ is not a Galilean one due to the impossibility to find a
 set of
points at rest on it,we cannot place the particle exactly on the
 origin
 $O^{\prime}$,since there would be no exact location on $x^{\prime}=0$ ($O^{\prime}$)
 (an uncertainty $\Delta x^{\prime}=\overline{O^{\prime}C}$: see Figure 3).
  Actually we want to show that $\Delta x^{\prime}$ (Fig.3) is a
 function
which should depend on speed $v$ of $S^{\prime}$ with respect to
 $S_V$,namely,for example,if $S^{\prime}\rightarrow S_V$ ($v\rightarrow
 V$),then we should have $\Delta x^{\prime}\rightarrow\infty$ (infinite
 uncertainty),which is due to the non-local aspect of
 the ultra-referential $S_V$. On the other hand,if
 $S^{\prime}\rightarrow S_c
(v\rightarrow c)$,then we should have $\Delta x^{\prime}\rightarrow 0$
 (much better located on $O^{\prime}$). Thus let us search for a function $\Delta
 x^{\prime}=\Delta x^{\prime}(v)=\Delta x^{\prime}_v$,starting from
Figure 3.\\

\begin{figure}
\includegraphics[scale=0.85]{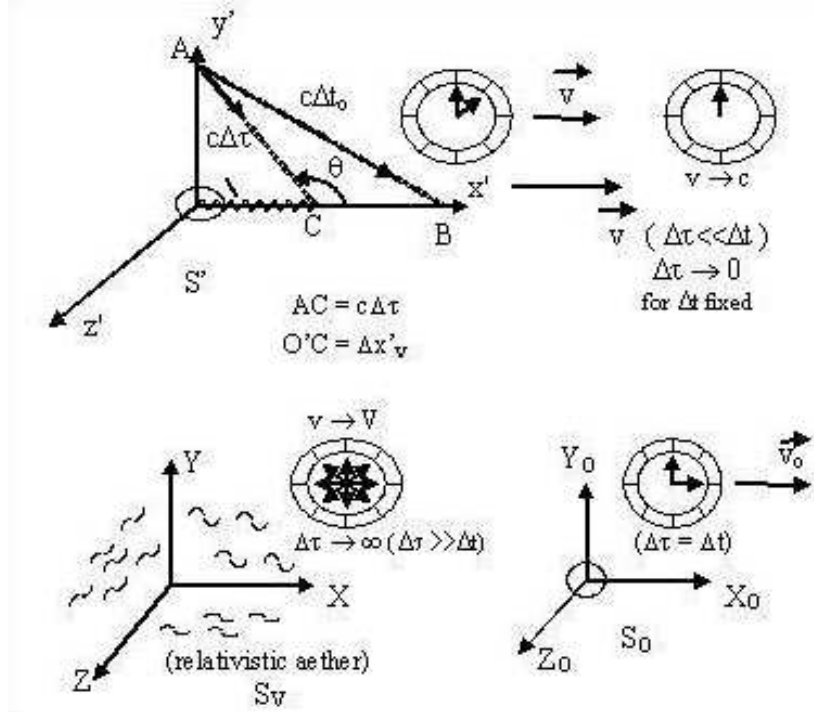}
\caption{\it We have four imaginary clocks associated with non-Galilean
 reference frames $S_0$, $S^{\prime}$, the ultra-referential $S_V$ (for
 $V$) and also $S_c$ (for $c$). We observe a new result,namely the
 proper time (interval $\Delta\tau$) elapses
much faster closer to infinite ($\Delta\tau\rightarrow\infty$) when one
approximates to $S_V$. On the other hand,it tends to stop
 ($\Delta\tau\rightarrow 0$) when $v\rightarrow c $,providing the
 strong
 symmetry for SSR. Here we are fixing $\Delta t~(\Delta(t_0))$ and
 letting
$\Delta\tau$ vary.}
\end{figure}

At the frame $S^{\prime}$ in Fig.3,let us consider that a photon is
 emitted from a point $A$ at $y^{\prime}$,in the direction $\overline
 {AO^{\prime}}$. This occurs only if $S^{\prime}$ were Galilean (at
 rest over
itself). However,since the electron cannot be thought of as a point at
 rest on
its proper non-Galilean frame $S^{\prime}$,and cannot be located
 exactly
 on $O^{\prime}$,its non-location $\overline {O^{\prime}C}$
 ($=\Delta x^{\prime}_v$)(see Fig.3)
 causes the photon to deviate from the direction $\overline
 {AO^{\prime}}$ to $\overline {AC}$. Hence,instead of just the
 segment
 $\overline {AO^{\prime}}$,a rectangular triangle $AO^{\prime}C$ is
 formed at the
proper non-Galilean reference frame $S^{\prime}$,where it is not
 possible
 to find a set of points at rest.

 From the non-Galilean frame $S_0$ (``quantum rest''),which plays the
 role of
 $S$,from where one ``observes'' the motion of $S^{\prime}$ across
 $S_V$,one can see the trajectory $\overline {AB}$ for the photon. Thus
 the
 rectangular
 triangle $AO^{\prime}B$ is formed. Since the vertical leg $\overline
 {AO^{\prime}}$ is common to the triangles $AO^{\prime}C$
 (for $S^{\prime}$) and
 $AO^{\prime}B$ (for $S_0\equiv S$),we have

 \begin{equation}
(\overline {AO^{\prime}})^2=(\overline {AC})^2-(\overline
 {O^{\prime}C})^2=
(\overline {AB})^2 - (\overline {O^{\prime}B})^2, 
\end{equation}

                             or else

 \begin{equation}
(c\Delta\tau)^2-(\Delta x^{\prime}_v)^2=
(c\Delta t_0)^2 - (v\Delta t_0)^2. 
\end{equation}
If $\Delta x^{\prime}(v)=\Delta x^{\prime}_v=0$ ($V\rightarrow
 0\Rightarrow
 S_V\equiv S_0(\equiv S)$),
we go back to the classical
 case (SR),where we consider for instance a train wagon
 ($S^{\prime}$),which is moving in relation to a fixed rail ($S$). At a point A on the
 ceiling
 of the wagon,there is a laser that releases photons toward
 $y^{\prime}$,reaching the point $O^{\prime}$ assumed in the origin of $S^{\prime}$
  (on the floor of the train wagon). For
 Galilean-$S^{\prime}$,the trajectory of the photon is $\overline
 {AO^{\prime}}$. For Galilean-$S$,its trajectory is $\overline {AB}$.

 Since $\Delta x^{\prime}_v$ is a function of $v$,assumed as a kind of
 ``displacement" (uncertainty)
given on the proper non-Galilean reference frame $S^{\prime}$,we may
write it in the following way:

\begin{equation}
(\Delta x^{\prime}_v)= f(v)\Delta\tau,
\end{equation}
where $f(v)$ is a
function of $v$,
 which also presents dimension of velocity,i.e.,it is a certain
velocity in
SSR,which could be thought of as a kind of
 {\it internal motion $v_{int}$} of the particle,being responsible for
 the increasing or dilation (stretch) of an internal dimension of the
 particle
 on its own
non-Galilean frame $S^{\prime}$. Such an internal dilation is given by
 the
 non-classical ``displacement"
$\Delta x^{\prime}_v=\overline{O^{\prime}C}$ (see Fig.3). This leads us
 to think
 that there is an uncertainty of position for the particle,as we will
 see later. Hence,substituting (47) into (46),we obtain

\begin{equation}
\Delta\tau[1-\frac{(f(v))^2}{c^2}]^{\frac{1}{2}}=
\Delta t(1-\frac{v^2}{c^2})^{\frac{1}{2}},
\end{equation}
where we use the notation
$\Delta t_0$ or $\Delta t$ $(S_0\equiv S)$,and where we have
 $f(v)=v_{int}$ to be duly interpreted.

Thus,since we have $v\leq c$,we should have $f(v)\leq c$ in order to
 avoid an imaginary number in the 1st. member of (48).

The domain of $f(v)$ is such that $V\leq v\leq c$. Thus,let us also
 think that its image is $V\leq f(v)\leq c$,since $f(v)$ has dimension
 of velocity
 and also represents a speed $v_{int}$ (internal motion),
which also must be limited for the extremities $V$ and $c$.

Let us make $[f(v)]^2/c^2 = f^2/c^2 =v_{int}^2/c^2=\alpha^2$,
 whereas we already know that
 $v^2/c^2 =\beta^2$. $v$ is the well-known external motion
(spatial velocity). Thus we have the following cases originated from
 (48),namely:

- (i) When $v\rightarrow c$ ($\beta\rightarrow\beta_{max}=1$),the
 relativistic
 correction in its 2nd. member (right-hand side) prevails,whereas the
 correction
 on the left-hand side becomes practically neglected,i.e.,we
 should have
 $v_{int}=f(v)<<c$,where $lim_{v\rightarrow
 c}f(v)=f_{min}=(v_{int})_{min}=V$
 ($\alpha\rightarrow\alpha_{min}=V/c=\xi$). $\xi\cong 3.57\times
 10^{-24}$ (refer to (41)).

 -(ii) On the other hand,due to idea of symmetry,if $v\rightarrow V$
 ($\beta\rightarrow\beta_{min}=V /c =\xi$),there is no substantial
 relativistic
 correction on the right-hand side of (48),whereas the correction on
 the
 left-hand side becomes now considerable,namely we should have
$lim_{v\rightarrow V}f(v)=f_{max}=(v_{int})_{max}
=c$ ($\alpha\rightarrow\alpha_{max}=1$).

 In short,from (i) and (ii),we observe that,if $v\rightarrow
 v_{max}=c$,
then $f\rightarrow f_{min}=(v_{int})_{min}=V$,and if $v\rightarrow
 v_{min}=V$,
 then $f\rightarrow f_{max}=(v_{int})_{max}=c$. So now we perceive
 that the internal motion $v_{int}$ ($=f(v)$) works like a
 reciprocal speed ($v_{Rec}$) in such a symmetrical structure of
 space-time in SSR. In other words,we notice that the
(external or spatial) velocity $v$
 increases to $c$ whereas the internal (reciprocal) one ($v_{int}=v_{Rec}$)
decreases to $V$. On the
other hand,when $v$ tends to $V$($S_V$),$v_{int}$ tends to $c$,leading
to a large internal stretch (uncertainty $\Delta x^{\prime}_v$) due to
a non-locality behavior much closer to the ultra-referential $S_V$.
 Due to this fact,we reason that

\begin{equation}
f(v)=v_{int}=v_{Rec}=\frac{a}{v},
\end{equation}
where $a$ is a constant that has dimension of square speed. Such
a reciprocal
velocity $v_{Rec}$ will be better understood later. It is interesting
to know that a similar idea of considering an internal motion for
microparticles was also thought by Natarajan\cite{26}.

In addition to (48) and (49),we already know that,at the referential
$S_0$ (see Fig.2 and Fig.3),we should
have the condition of equality of the time intervals,namely
$\Delta t=\Delta\tau$ for $v=v_0$,which,in accordance with (48),occurs
only if

\begin{equation}
\frac{[f(v_0)]^2}{c^2}=\frac{v_0^2}{c^2}\Leftrightarrow f(v_0)=v_0
\end{equation}

By comparing (50) with (49) for the case $v=v_0$,we obtain

\begin{equation}
a=v_0^2
\end{equation}

 Substituting (51) into (49),we obtain

\begin{equation}
f(v)=v_{int}= v_{Rec}=\frac{v_o^2}{v}
\end{equation}

According to (52) and also considering (i) and (ii),indeed we observe
respectively that $f(c)=V=v_0^2/c$ ($V$ is the reciprocal velocity of
 $c$) and
$f(V)=c=v_0^2/V$ ($c$ is the reciprocal velocity of $V$),from where
we immediately obtain

\begin{equation}
v_0=\sqrt{cV}
\end{equation}

As we already know the value of $V$ (refer to (43)) and $c$,we
obtain the
velocity of ``quantum rest'' $v_0\cong 5.65\times 10^{-4}m/s$,which is
also universal just because it depends on the universal constants $c$
 and $V$. However,we must stress that only $c$ and $V$ remain invariant under
speed transformations in such a space-time of SSR (section 6).

Finally,by substituting (53) into (52) and after into (48),we finally
obtain

\begin{equation} 
\Delta\tau\sqrt{1-\frac{V^2}{v^2}}=\Delta
 t\sqrt{1-\frac{v^2}{c^2}},
\end{equation}
where $\alpha=f(v)/c=v_{int}/c=
V/v$ and $\beta=v/c$ inside (54). In fact,if $v=v_0=\sqrt{cV}$ in
 (54),so
 we have $\Delta\tau =\Delta t$. Therefore we conclude that $S_0~(v_0)$
 is the intermediate (non-Galilean) reference frame such that,if:

a) $v>>v_0$ ($v\rightarrow c$)$\Rightarrow\Delta t>>\Delta\tau$: It is
 the well-known {\it time dilation \/}.

b) $v<<v_0$ ($v\rightarrow V$) $\Rightarrow\Delta t<<\Delta\tau$: Let
 us call this new result {\it contraction of time \/}. This shows us
 the
 novelty that
 the proper time interval ($\Delta\tau$) is variable,so that it
 may expand in relation to the improper
 one ($\Delta t$ in $S_0$). $\Delta\tau$ is an intrinsic variable for
 the
 particle on its
 proper non-Galilean frame $S^{\prime}$. Such an effect of dilation
 of $\Delta\tau$ with respect to $\Delta t$ would become more evident
 only for $v\rightarrow V~(S_V)$,since we would have
 $\Delta\tau\rightarrow\infty$
in such a limit $S_V$. In other
 words,this means that the proper time ($S^{\prime}$) would elapse
 much faster than the improper one at $S_0$.

  In SSR,it is interesting to notice that we
 restore the newtonian regime when $V<<v<<c$,which represents
 an intermediate regime of speeds,where we can make the approximation
 $\Delta\tau\approx\Delta t$.

  Substituting (52) into (47) and also considering (53),we obtain

\begin{equation}
\overline{O^{\prime}C}=\Delta x^{\prime}_v=v_{int}\Delta\tau=
v_{Rec}\Delta\tau=\frac{v_0^2}{v}\Delta\tau=
\frac{\it{V}c}{v}\Delta\tau=\alpha c\Delta\tau
\end{equation}

 Actually we can verify that,if $V\rightarrow 0$  or
$v_0\rightarrow 0$,this implies $\overline {O^{\prime}C}=\Delta
 x^{\prime}_v=0$,
 restoring the classical case (SR),where there is no such an internal motion.
 And also,if $v>>v_0$,this implies $\Delta x^{\prime}_v\approx 0$,i.e.,we have
an approximation where the internal motion is neglected.

From (55),it is important to observe that,if $v\rightarrow c$,we have
 $\Delta x^{\prime}(c)=V\Delta\tau$ and,if $v\rightarrow V$ ($S_V$),
 we have
 $\Delta x^{\prime}(V)=c\Delta\tau$. This means that,when the particle
 momentum with respect to $S_V$
 increases ($v\rightarrow c$),it becomes much more localized upon itself
 over $O^{\prime}$ ($V\Delta\tau\rightarrow 0$) and,when its momentum decreases
 ($v\rightarrow V$),it becomes much less localized over $O^{\prime}$,because it
 gets much closer to
 the non-local ultra-referential $S_V$,where $\Delta
 x^{\prime}_v=\Delta
 x^{\prime}_{max}=\overline{0^{\prime} C}_{max}=
c\Delta\tau\rightarrow\infty$. Thus,now we begin to
 perceive
that the velocity $v$ (momentum) and the position (non-localization $\Delta
 x^{\prime}_v
=v_{Rec}\Delta\tau$) operate like
 mutually reciprocal quantities in such a space-time of SSR,since
 the non-localization is $\Delta x^{\prime}_v\propto v_{Rec}\propto v^{-1}$
 (see (49) or (52)). This really provides a basis for the fundamental
 comprehension of the quantum uncertainties in a context of objective reality
 of the space-time,according to Einstein's vision\cite{18}.

It is very interesting to observe that we may write $\Delta
 x^{\prime}_v$ in the following way:

\begin{equation}
\Delta x^{\prime}_v = \frac{(V\Delta\tau)(c\Delta\tau)}{v\Delta\tau}
\equiv\frac{\Delta x^{\prime}_5\Delta x^{\prime}_4}{\Delta
 x^{\prime}_1},
\end{equation}
where $V\Delta\tau=\Delta x^{\prime}_5$, $c\Delta\tau=\Delta
 x^{\prime}_4$ and
$v\Delta\tau=\Delta x^{\prime}_1$. We also know that $c\Delta t_0\equiv
c\Delta t=\Delta x_4$ and $v\Delta t_0\equiv v\Delta t=\Delta x_1$ for
the frame $S(\equiv S_0)$. So we write (46) in the following way:

\begin{equation} 
\Delta x^{\prime 2}_4-
 \frac{\Delta x^{\prime 2}_5\Delta x^{\prime 2}_4}{\Delta x^{\prime
 2}_1}=\Delta x_4^2 -\Delta x_1^2,
\end{equation}
where $\Delta x^{\prime}_5$ corresponds to a fifth dimension of
temporal nature.
 Therefore we may already conclude that the new geometry of space-time
 has three spatial dimensions ($x_1$, $x_2$, $x_3$) plus two temporal
 dimensions
 ($c\Delta t$, $V\Delta\tau$),being $V\Delta\tau$ normally hidden.
  However,we
will perceive elsewhere that we can also describe such a space-time in a
 compact form as effectively a 4-dimensional structure,because
 $V\Delta\tau$
and $c\Delta t$ represents two complementary aspects of the same
temporal nature,
and also mainly because $V\Delta\tau$ appears as an implicit variable
for the space-time interval $c\Delta\tau$ (see (61), (62) or (63)).

 If $\Delta x^{\prime}_5\rightarrow 0$ ($V\rightarrow 0$),we restore
 the invariance of the 4-dimensional interval in Minkowski space as a
 particular
 case,that is,$\Delta S^2=\Delta x_4^2-\Delta x_1^2=
\Delta S^{\prime 2}=\Delta x^{\prime 2}_4$.

As we have $\Delta x^{\prime}_v>0$,we observe that
$\Delta S^{\prime 2}=\Delta x^{\prime 2}_4>\Delta S^2=
\Delta x_4^2-\Delta x_1^2$. Hence,we may write (57),as follows:

\begin{equation}
\Delta S^{\prime 2} = \Delta S^2 + \Delta x^{\prime 2}_v,
\end{equation}
where $\Delta S^{\prime}=\overline {AC}$, $\Delta x^{\prime}_v=
\overline {O^{\prime}C}$ and $\Delta S=\overline {AO^{\prime}}$ 
(refer to Fig.3).

For $v>>V$ or also $v\rightarrow c$,we have $\Delta
 S^{\prime}\approx\Delta S$,
 hence $\theta\approx\frac{\pi}{2}$ (see Fig.3). In macroscopic
 world
 (or very large masses),we have $\Delta x^{\prime}_v=\Delta x^{\prime}_5=0$
 (hidden dimension),hence $\theta=\frac{\pi}{2}\Rightarrow\Delta
 S^{\prime}=\Delta S$. The quantum uncertainties can be neglected in
 such a particular regime (Galilean reference frames of SR).

   For $v\rightarrow V$,we would have $\Delta S^{\prime}>>\Delta S$,
 where
 $\Delta S^{\prime}\approx c\Delta\tau$,with
 $\Delta\tau\rightarrow\infty$ and
 $\theta\rightarrow\pi$. In this new relativistic limit (relativistic
 ``ether" $S_V$),due to the maximum non-localization
$\Delta x^{\prime}_v\rightarrow\infty$,the 4-dimensional interval
 $\Delta S^{\prime}$ loses completely its equivalence in respect to
 $\Delta S$,because 5th dimension ($V\Delta\tau$) increases drastically
 much closer to such a limit,i.e.,$\Delta x^{\prime}_5\rightarrow\infty$. So it
ceases to be hidden for such very special case.

   Equation (58) or (57) shows us a break of the $4$-interval
 invariance
($\Delta S^{\prime}\neq\Delta S$),which becomes noticeable only at the
 limit
 $v\rightarrow V$ ($S_V$). However,a new invariance is restored when
 we implement a 5th.dimension ($x^{\prime}_5$) to be intrinsic to the particle
 (frame $S^{\prime}$) through the definition of a new (effective)
 general interval,where the interval $V\Delta\tau$ appears as an implicit
 variable,namely:

\begin{equation}
\Delta S_5=\sqrt{\Delta S^{\prime 2}-\Delta x^{\prime 2}_v}=
\Delta x^{\prime}_4\sqrt{1-\frac{\Delta x^2_5}{\Delta x^{\prime 2}_1}}=
c\Delta\tau\sqrt{1-\frac{V^2}{v^2}},
\end{equation}
such that $\Delta S_5\equiv\Delta S$ (see (58)).

We have omitted the index $\prime$ for $\Delta x_5$,as such an interval
is given
only at the non-Galilean proper reference frame ($S^{\prime}$),that is
intrinsic to the
particle. Actually such a 5-interval or simply an effective 4-interval
$c\Delta\tau*=c\Delta\tau\sqrt{1-\alpha^2}$ guarantees the existence of
a certain effective internal dimension for the electron. However,from a
practical viewpoint,for experiments of higher energies,the electron
approximates more and more to a punctual particle,since $\Delta x_5$
becomes hidden. So in order to detect its internal dimension,
it should be at very low energies,namely very close to $S_V$.

Comparing (59) with the left side of equation (54),we may
alternatively write

\begin{equation}
\Delta t=\Psi\Delta\tau=\frac{\Delta S_5}{c\sqrt{1-\frac{v^2}{c^2}}}=
\Delta\tau\frac{\sqrt{1-\frac{V^2}{v^2}}}{\sqrt{1-\frac{v^2}{c^2}}} , 
\end{equation}
where $\Delta S_5$ is the invariant effective interval given at the
frame $S^{\prime}$. We have $\Psi=\frac{\sqrt{1-\alpha^2}}{\sqrt{1-\beta^2}}
=\frac{\sqrt{1-\frac{V^2}{v^2}}}{\sqrt{1-\frac{v^2}{c^2}}}$ and,
 alternatively,we can also write $\Psi=\frac{\sqrt{1-\beta^2_{int}}}
{\sqrt{1-\alpha^2_{int}}}=\frac{\sqrt{1-\frac{v_{int}^2}{c^2}}}
{\sqrt{1-\frac{V^2}{v_{int}^2}}}$,since
 $\alpha=V/v=\beta_{int}=v_{int}/c$
and $\beta=v/c=\alpha_{int}=V/v_{int}$,from where we get
 $v_{int}=v_{Rec}=cV/v=v_0^2/v$ (see (52)). Only for $v=v_0$,we obtain $v_{int}=v=v_0$.
 Although we cannot obtain directly $v_{int}$ by any experiment (just
the uncertainty $\Delta x$ is obtained),we could also use
$\Psi$ in its alternative form $\Psi(v_{int})$. However,let us use $\Psi(v)$.

 For $v>>V$,we get $\Delta t\approx\gamma\Delta\tau$,where
$\Psi\approx\gamma=(1-\beta^2)^{-1/2}$.

 Substituting (55) into (46) and using the notation $\Delta
 t_0\equiv\Delta t$,we obtain

\begin{equation}
c^2\Delta\tau^2=\frac{1}{(1-\frac{V^2}{v^2})}[c^2\Delta t^2-v^2\Delta
 t^2],
\end{equation}
from where,we also obtain the equation (54).

By placing (61) in a differential form and manipulating it,we will
obtain

\begin{equation}
c^2(1-\frac{V^2}{v^2})\frac{d\tau^2}{dt^2} + v^2 = c^2
\end{equation}

We may write (62) in the following alternative way:

\begin{equation}
\frac{dS_5^2}{dt^2} + v^2 = c^2,
\end{equation}
where $dS_5=c\sqrt{1-\frac{V^2}{v^2}}d\tau$.

 Equation (62) shows us that the speed related to the marching of time
 (``temporal-speed''),which is
 $v_t=c\sqrt{1-\frac{V^2}{v^2}}\frac{d\tau}{dt}$,
 and the spatial speed,which is $v$ in relation to the background
 field for $S_V$ form respectively the vertical and horizontal legs of a
rectangular triangle.

 We have $c=(v_t^2+v^2)^{1/2}$,which represents
 the space-temporal velocity of any particle (hypothenuse of the
 triangle=$c$). The
 novelty here is that such a space-time implements the
 ultra-referential $S_V$. Such an implementation arises at the vertical
 leg $v_t$ of such a rectangular triangle.

 We should consider 3 importants cases as follow:

 a)    If $v\approx c$,then $v_t\approx 0$,where $\Psi>>1$,since
 $\Delta t>>\Delta\tau$ ({\it dilation of time\/}).

 b)    If $v=v_0=\sqrt{cV}$,then $v_t=\sqrt{c^2-v_0^2}$,where
 $\Psi=\Psi_0=\Psi(v_0)=1$,
 since $\Delta t=\Delta\tau$ ({\it ``quantum rest'' $S_0$\/}).

 c)    If $v\approx V$,then $v_t\approx\sqrt{c^2-V^2}
 =c\sqrt{1-\xi^2}$,where
$\Psi<<1$,since $\Delta t<<\Delta\tau$ ({\it contraction of time \/}).

  In SR,when $v=0$,we have $v_t=v_{tmax}=c$. However,in accordance
 with SSR,
 due to the existence of a minimum limit $V$ of spatial speed for the
 horizontal
 leg of the triangle,we see that the maximum temporal-speed is
 $v_{tmax}<c$.
  This means that we have $v_{tmax}=c\sqrt{1-\xi^2}$. Such a result introduces a
 strong
 symmetry in such a space-time of SSR,in the sense that
 both of
 spatial and temporal speeds $c$ become unattainable for all massive
 particles.

 The speed $v=c$ is represented by the photon (massless particle),
 whereas $v=V$ is definitely inaccessible for any particle. Actually
we have $V<v\leq c$,but,in this sense,we have a certain asymmetry,as there
is no particle at the ultra-referential $S_V$ where there should be a
kind of {\it sui generis \/} vacuum energy density ($\rho^{(2)}$) to be studied
elsewhere.

  In order to produce a geometric representation for that problem
 ($V<v\leq c$),
 let us assume the world line of a particle limited by the surfaces of
 two cones,as shown in Figure 4.

\begin{figure}
\begin{center}
\includegraphics[scale=0.9]{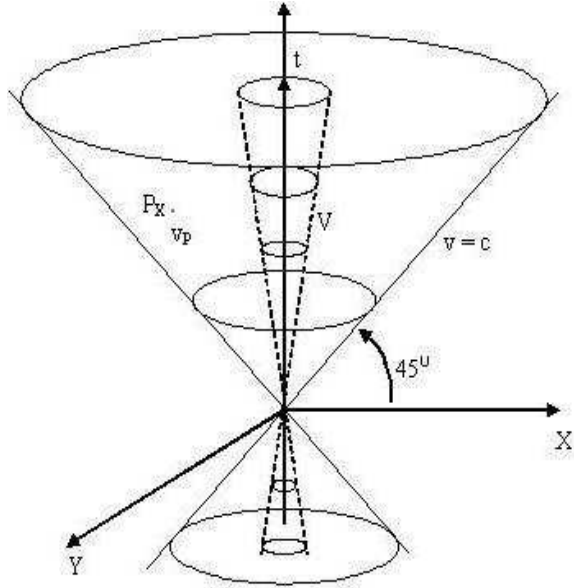}
\end{center}
\caption{\it The external and internal conical surfaces represent
 respectively $c$ and $V$,where $V$ is represented by the dashed line,
 that is a definitely prohibited boundary. For a point $P$ in the interior
 of the two conical surfaces,there is a corresponding internal conical
surface,such that $V<v_P\leq c$.}
\end{figure}

 A spatial speed $v=v_{P}$ in the representation of light cone
shown in
Figure 4 (horizontal leg of the rectangular triangle) is associated
with a
temporal speed $v_t=v_{tP}=\sqrt{c^2-v_P^2}$ (vertical leg of the same
triangle)
given in another cone representation,which could be denominated
 {\it temporal cone} (Figure 5).

\begin{figure}
\begin{center}
\includegraphics[scale=0.7]{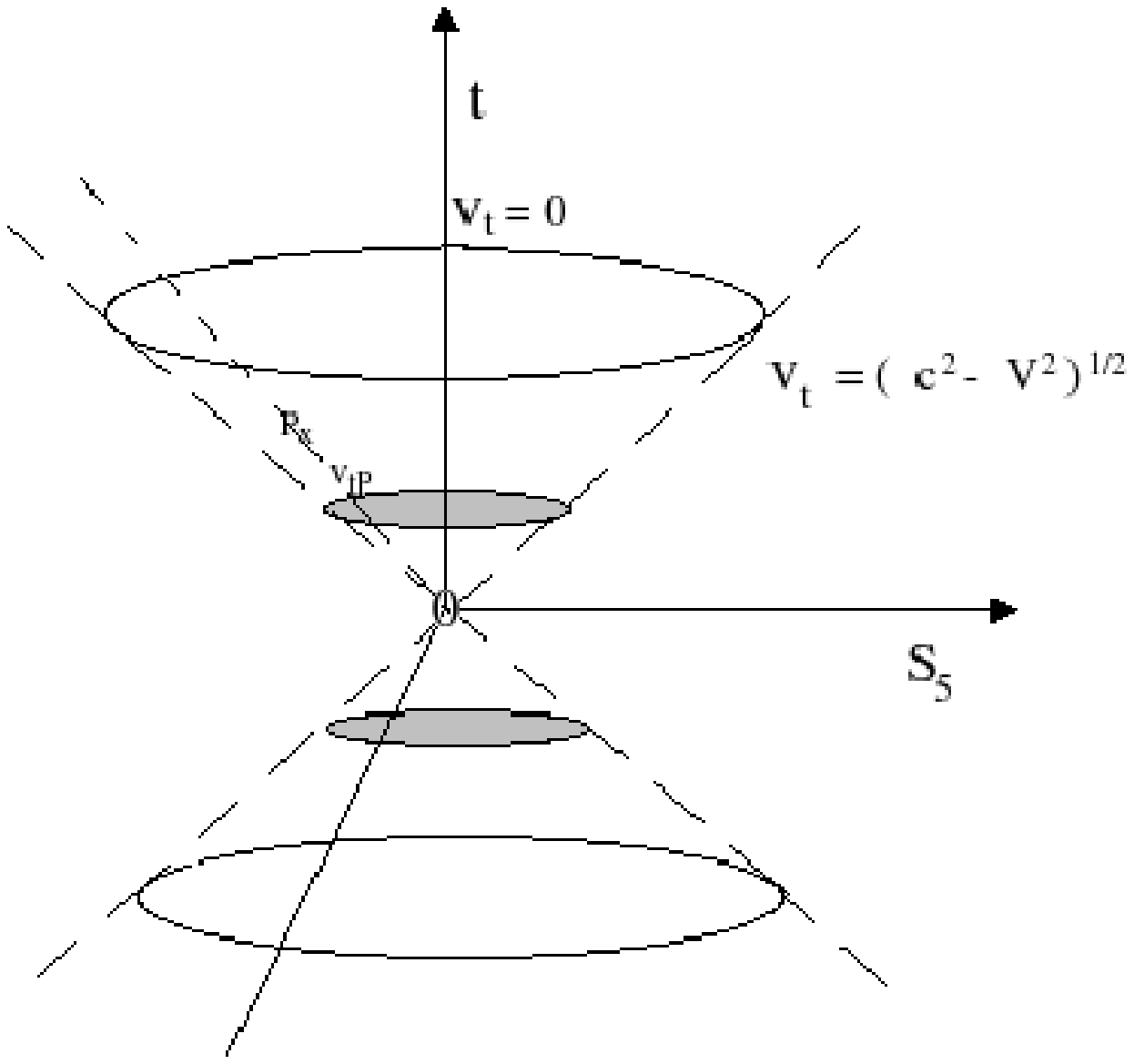}
\end{center}
\caption{\it Comparing this Figure 5 with Figure 4,we notice that the
 dashed line
 on the internal cone of Figure 4 ($v=V$) corresponds to the dashed
 line on the
 surface of the external cone of this Figure 5,where
 $v_t=\sqrt{c^2-V^2}$,which
 represents a definitely forbidden boundary in this cone representation
 of temporal speed $v_t$. On the other hand,$v=c$ (photon) is represented
 by the solid line of Figure 4,which corresponds to the temporal speed
 $v_t=0$ in this Figure 5,coinciding with the vertical axis $t$. In short,we always have
 $v^2 + v_t^2=c^2$,being $v$ for spatial
 (light) cone (Figure 4) and $v_t$ for temporal cone represented in
 this Figure 5,such that an internal point $P_x$ is related to a temporal velocity $v_{tP}$,where
 $0~(photon)\leq v_{tP}(=\sqrt{c^2-v_P^2})<\sqrt{c^2-V^2}$. The horizontal axis is
 $S_5=c\sqrt{1-V^2/v^2}\tau$,so that $v_t=dS_5/dt=c\sqrt{1-V^2/v^2}d\tau/dt=\sqrt{c^2-v^2}$
 (see equation (54)).}
\end{figure}

   We must observe that a particle moving just at one spatial dimension
always goes only to left or to right,since the unattainable non-null minimum limit of
speed $V$ forbids it
to stop its spatial velocity ($v=0$) in order to return at this same spatial dimension.
 On the other hand,in a complementary way to $V$,
the limit $c$ is temporal in the sense that it forbids to stop the time (temporal velocity $v_t=0$)
and also to come back to the past. However,if we consider more than one spatial
dimension,at least 2 spatial dimensions($xy$),
the particle can now return by moving at the additional
dimension(s). So SSR provides the reason why we must have more than one (1) spatial
dimension ($d>1$) for representing movement in reality,although we could consider $1d$ just as
a good approximation for some cases in classical space-time of SR (classical
objects). Such a minimum limit $V$ has deep implications for understanding the
irreversible aspect of time connected to movement,since we can now distinguish
the motions to left and to right in the time. Such an asymmetry generated by SSR really
deserves a deeper treatment elsewhere.

 Based on the relation (61) or also by substituting (55) into (46),we obtain

  \begin{equation}
  c^2\Delta t^2 - v^2\Delta t^2 = c^2\Delta\tau^2 -
 \frac{v_0^4}{v^2}\Delta\tau^2
  \end{equation}

 In (64),when we transpose the 2nd.term from the left side to
 the right side and divide the equation by $\Delta t^2$,we obtain
 (62) in differential form. Now,it is important to observe that,upon
 transposing the 2nd.term from the right side to the left one
 and dividing the equation by $\Delta\tau^2$,we obtain the following equation
 in the differential form,namely:

  \begin{equation}
  c^2(1-\frac{v^2}{c^2})\frac{dt^2}{d\tau^2} + \frac{v_0^4}{v^2}=c^2
  \end{equation}

 From (59) and (54),we obtain
 $dS_5=cd\tau\sqrt{1-\alpha^2}=cdt\sqrt{1-\beta^2}$. Hence
 we can write (65) in the following alternative way:
  \begin{equation}
  \frac{dS_5^2}{d\tau^2} + \frac{v_0^4}{v^2}= c^2
  \end{equation}

 We see that equation (65) or (66) reveals a complementary way of
 viewing equation (62) or (63). This leads us to that idea of reciprocal space
 for conjugate quantities. Thus let us write (65) or (66) in the following way:

  \begin{equation}
  v_{tRec}^2 + v_{Rec}^2 =c^2,
  \end{equation}
where $v_{tRec}=(v_t)_{int}=dS_5/d\tau=
c\sqrt{1-\frac{v^2}{c^2}}\frac{dt}{d\tau}$,which represents an {\it internal
(reciprocal) temporal velocity}. The internal (reciprocal) spatial velocity
is $v_{int}=v_{Rec}=f(v)=\frac{v_0^2}{v}$. Therefore we can also represent a
rectangular triangle,but now displayed in a reciprocal space. For example,
if we assume $v\rightarrow c$ (equation (62)),we obtain
 $v_{Rec}=lim_{v\rightarrow c}f(v)\rightarrow\frac{v_0^2}{c}=V$ (equation
 (65)). In this same case,we have $v_t\rightarrow 0$ (equation (62)) and
$v_{tRec}=\frac{dS_5}{d\tau}\rightarrow\sqrt{c^2-V^2}$ (equation (65) or (66)).
 On the other hand,if $v\rightarrow V$ (eq.(62)),we have
 $v_{Rec}\rightarrow
\frac{v_0^2}{V}=c$ (eq.(65)),where $v_t\rightarrow\sqrt{c^2-V^2}$
 (eq.(62))
 and $(v_t)_{int}=v_{tRec}\rightarrow 0$ (eq.(65)). Thus we should
 observe that
 there are altogether four cone representations in such a symmetrical
 structure of space-time in SSR,namely:
 
\begin{equation}
 two~spatial~representations:\left\{\begin{array}{ll}
a_1) v=\frac{dx}{dt}, in~ equation~ (62), \\
represented~in~Fig.4; \\
 b_1) v_{Rec}=\frac{d x^{\prime}_v}{d\tau}=\frac{v_0^2}{v}, in~
 equation~(65).
\end{array}
 \right.
 \end{equation}

\begin{equation}
 two~temporal~representations:\left\{\begin{array}{ll}
a_2) v_t=\frac{dS_5}{dt}=c\sqrt{1-\frac{V^2}{v^2}}\frac{d\tau}{dt}=c\sqrt{1-\frac{v^2}{c^2}},\\
 in~equation~(62),represented~in~Fig.5;\\

b_2) v_{tRec}=\frac{dS_5}{d\tau}=c\sqrt{1-\frac{v^2}{c^2}}\frac{dt}{d\tau}=
c\sqrt{1-\frac{V^2}{v^2}},\\
in~ equation ~(65).
\end{array}
 \right.
 \end{equation}

The chart given in Figure 6 shows us those four representations.
\begin{figure}
\begin{center}
\includegraphics[scale=0.72]{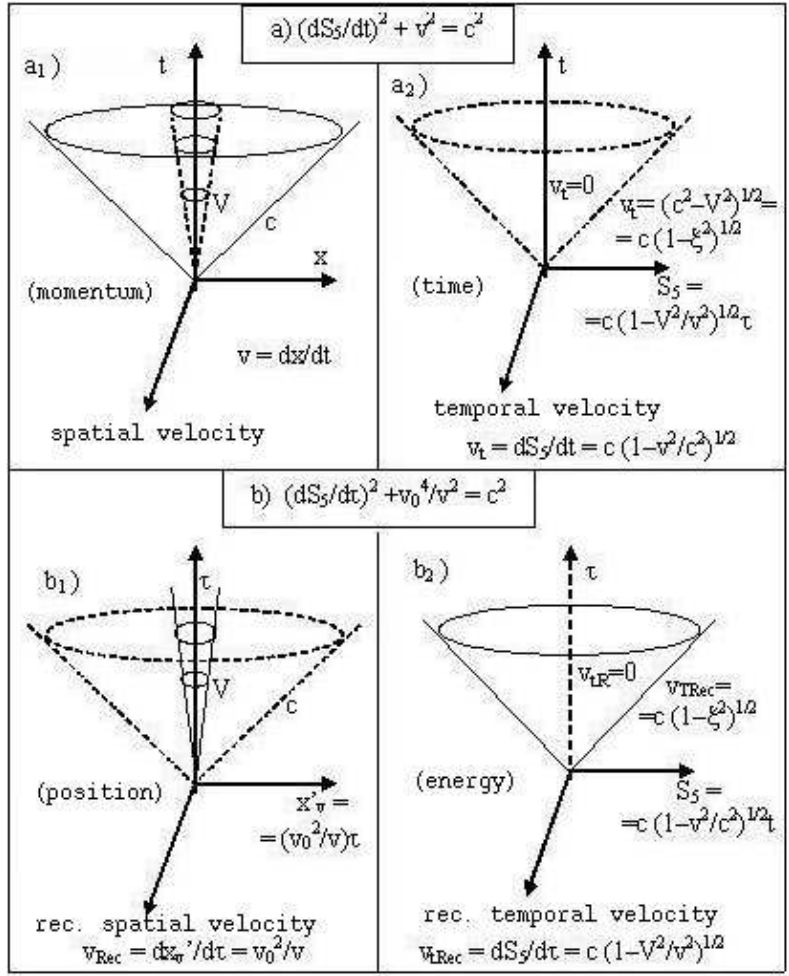}
\end{center}
\caption{\it The spatial representations in $a_1$ (also shown in Figure 4)
and $b_1$ are related
 respectively to velocity $v$ (momentum) and position (non-localization
 $\Delta
 x^{\prime}_v=f(v)\Delta\tau=v_{int}\Delta\tau=
v_{Rec}\Delta\tau=(v_0^2/v)\Delta\tau$),which
 represent conjugate (reciprocal) quantities in space. On the other
 hand,the
 temporal representations in $a_2$ (also shown in Figure 5) and
 $b_2$ are related respectively to time
 ($\propto v_t$) and energy ($\propto v_{tRec}=(v_t)_{int}
\propto v_t^{-1}$),which
represent conjugate (reciprocal) quantities in the time. Hence we can
perceive that such four cone representations of SSR provide a basis for the
fundamental understanding of the two uncertainty relations.}
\end{figure}

Now,by considering (54),(60),(69) and also looking at $a_2$ and $b_2$
in Fig.6,we may observe that
 \begin{equation}
 \Psi^{-1}=\frac{\Delta\tau}{\Delta t}=
 \frac{\sqrt{1-\frac{v^2}{c^2}}}{\sqrt{1-\frac{V^2}{v^2}}}=
\frac{v_t}{c\sqrt{1-\frac{V^2}{v^2}}}= \frac{v_t}{v_{tRec}}\propto
 (time)
  \end{equation}

  and

  \begin{equation}
\Psi=\frac{\Delta t}{\Delta\tau}=
 \frac{\sqrt{1-\frac{V^2}{v^2}}}{\sqrt{1-\frac{v^2}{c^2}}}=
\frac{v_{tRec}}{c\sqrt{1-\frac{v^2}{c^2}}}=\frac{v_{tRec}}{v_t}\propto
 E
~(Energy\propto (time)^{-1}) 
  \end{equation}

 From (71),since we have energy $E\propto\Psi$,we write $E=E_0\Psi$,where
 $E_0$ is a constant of proportionality. Hence,if we consider $E_0=m_0c^2$,we
 obtain

\begin{equation}
E= m_0c^2\frac{\sqrt{1-\frac{V^2}{v^2}}}{\sqrt{1-\frac{v^2}{c^2}}}, 
\end{equation}
where  $E$ is the total energy of the particle in relation to the
absolute inertial frame of universal background field $S_V$. Such a result
shall be explored in a coming article about the dynamics of the
particles in SSR. In (71) and (72),we observe that,if $v\rightarrow c\Rightarrow
 E\rightarrow\infty$ and $\Delta\tau\rightarrow 0$ for $\Delta t$
 fixed. If $v\rightarrow V\Rightarrow E\rightarrow 0$ and
 $\Delta\tau\rightarrow\infty$,
 also for $\Delta t$ fixed. If $v=v_0=\sqrt{cV}\Rightarrow E=E_0=m_0c^2$ (energy
 of ``quantum ``rest''''). Figure 7 shows us the graph for the energy $E$ in (72).

\begin{figure}
\begin{center}
\includegraphics[scale=0.4]{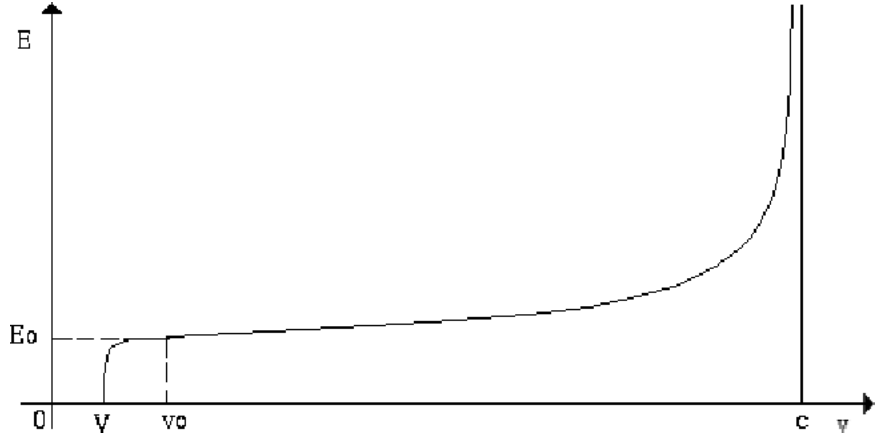}
\end{center}
\caption{\it $v_0$ represents the velocity of ``quantum rest'' in SSR,
 from where we get $E=E_0=m_0c^2$,being $\Psi_0=\Psi(v_0)=1$.}
\end{figure}

\subsection{The Uncertainty Principle}

The particle actual momentum (in relation to $S_V$) is $P=\Psi m_0 v$,
 whose conjugate value is $\Delta x^{\prime}_v=
 \frac{v_0^2}{v}\Delta\tau=\frac{v_0^2}{v}\Delta t\Psi^{-1}$,where
 $\Delta\tau=\Psi^{-1}\Delta t$ (refer to (54)). From $S_V$
 it would be possible to know exactly the actual momentum $P$ and the
 total energy $E$ of the particle,however,since $S_V$ represents an
 ultra-referential which is unattainable (non-local) and also
 inaccessible for
 us,so one becomes impossible to measure such quantities with
 accuracy. And for
 this reason,as a classical observer (local and macroscopic) is always
 at rest
 ($v=0$) in his proper reference frame $S$,he measures and interprets $E$
 without
 accuracy because his frame is Galilean,{\it being related
 essentially to
 macroscopic systems (a set of points at rest)},whereas on the other
 hand,non-Galilean reference frames for representing subatomic world in SSR
 are really always moving for any transformation in such a space-time
 {\it and
 therefore cannot be related to a set of points at rest}. Due to this
 {\it conceptual discrepancy between the nature of non-Galilean
 reference frames
 in SSR (no rest) and the nature of Galilean reference frames in SR for
 classical observes (with rest)},the total energy $E$ in SSR
 (eq.(72)) behaves
 as an uncertainty $\Delta E$ for such classical observers at rest,
 i.e.,$E$ (for
 $S_V$) $\equiv\Delta E$ (for any Galilean-$S$ at rest). Similarly
 $P$ also
behaves as an uncertainty $\Delta p$ ($P~(S_V)\equiv\Delta p$ (Galilean-S)) and,
in addition,the non-localization $\Delta x^{\prime}_v$ as simply an uncertainty
$\Delta x$. Hence we have

\begin{equation}
\Delta x^{\prime}_v P\equiv(\Delta x\Delta p)_{classical~ observer~S}=
\frac{v_0^2}{v}\Delta t\Psi^{-1}\Psi m_0 v= (m_0v_0)(v_0\Delta t)
\end{equation}

and

\begin{equation}
 \Delta\tau E\equiv(\Delta\tau\Delta E)_{classical~ observer~S}
=\Delta t\Psi^{-1}\Psi m_0 c^2=(m_0 c)(c\Delta t),
 \end{equation}
where we consider again $\Delta t$ fixed and let $\Delta\tau$
vary for each case. In obtaining (73) and (74),we also have considered the
relations $\Delta x^{\prime}_v=\frac{v_0^2}{v}\Delta\tau$,
$\Delta\tau=\Delta t\Psi^{-1}$, $P=\Psi m_0 v$ and $E=\Psi m_0c^2$.

Since we know the actual momentum $P$ of the particle moving across
the relativistic ``ether"-$S_V$,its de-Broglie wavelength is

 \begin{equation}
 \lambda=\frac{h}{P}=\frac{h}{\Psi m_0 v}=
\frac{h}{m_0
 v}\frac{\sqrt{1-\frac{v^2}{c^2}}}{\sqrt{1-\frac{V^2}{v^2}}}
 \end{equation}

If $v\rightarrow c$ $\Rightarrow\lambda\rightarrow 0$ ({\it spatial
 contraction \/} or {\it temporal dilation \/}),and if $v\rightarrow V$
 $\Rightarrow\lambda\rightarrow\infty$ ({\it spatial dilation \/} or
{\it temporal contraction\/}). In such a space-time of
SSR,actually we should interpret the spatial scales as wavelengths
$\lambda$ given at the background frame $S_V$,in accordance with (75).

  The relationship (75) shows us a strong symmetry that enables us to
 understand the
 space as an elastic structure,which is capable of contracting
 ($\lambda\rightarrow 0$ for $v\rightarrow c$) and also expanding
 ($\lambda\rightarrow\infty$ for $v\rightarrow V$ ($S_V$)).

 The wavelength $\lambda$ in (75) may be thought of as being related to
 the non-localization $\Delta x^{\prime}_v$,namely $\lambda\propto\Delta
 x^{\prime}_v$. Such a proportionality is verified by comparing (55) with
 (75) and also by considering $\Delta\tau=\Psi^{-1}\Delta t$. Hence we have

  \begin{equation} 
 \lambda\propto\Delta x^{\prime}_v=\frac{v_0^2}{v}\Delta\tau=
\frac{v_0^2}{v}\Delta t\frac{\sqrt{1-\frac{v^2}{c^2}}}
{\sqrt{1-\frac{V^2}{v^2}}},
  \end{equation}
where $\lambda\propto\Delta x^{\prime}_v$ ($\equiv\Delta
 x$)$=v_{int}\Delta\tau=v_{Rec}\Delta\tau\propto (v\Psi)^{-1}$.
 We also make $\Delta t$ fixed and let
 $\Delta\tau$ vary,such that $0<\Delta\tau<\infty$. Now,we can perceive
 that the quantum nature of the wave is derived from the internal
 motion $v_{int}=v_{Rec}$ of the proper particle,since its wavelength for $S_V$ is
 $\lambda\propto v_{Rec}$. This leads to a fundamental explanation for the
wave-particle duality in such a space-time of SSR. Natarajan\cite{26}
also used a kind of internal motion $v_{in}$\cite{26} of the microparticle to explain in
alternative way such a dual aspect of the matter. In approximation for
SR,we have $V\rightarrow 0$ (or also $v_0\rightarrow 0$),so that
 $v_{Rec}=0\Rightarrow\lambda=0$. Indeed this means that the wave nature of the
 matter is not included in SR.

 Now let us observe that,if we make $v=v_0$ in (76) and (75),and then compare
 these two results,we obtain

  \begin{equation}
  v_0\Delta t\equiv v_0T_0\sim\lambda_0=\frac{h}{m_0v_0}\sim 1m , 
  \end{equation}
where we fix $\Delta t\equiv T_0\sim\frac{h}{m_0v_0^2}$,$m_0$ being
the electron mass. $T_0$ represents the period of the wave with length
$\lambda_0$,such that $T_0\sim 10^3 s$. $\lambda_0$ is a special
standard intermediate scale for the frame $S_0$. Since $\lambda_0\sim 1m$,
indeed it represents a typical scale of a classical observer (human scale).

 Finally,by substituting (77) into (73),we obtain

  \begin{equation}
  \Delta x^{\prime}_v P\equiv\Delta x\Delta p\sim m_0v_0\lambda_0=h
  \end{equation}

  Now,it is easy to conclude that

  \begin{equation}
  \Delta\tau E\equiv\Delta\tau\Delta E\sim m_0 c\lambda_c=h,
  \end{equation}
 where $c\Delta t\equiv cT_c\sim\lambda_c=\frac{h}{m_0c}$ (refer to (74)).
  $\lambda_c\sim 10^{-12}m$ (Compton wavelength for the photon,whose
 energy $mc^2$ ($\propto e_sb_s$) must be equivalent to the electron
 energy $m_0c^2$ ($\propto e_{s0}b_{s0}$),that is,$m\equiv m_0$. In this
 instance,$\Delta t\equiv T_c\sim\frac{h}{m_0c^2}$).

 It is interesting to notice that
 $\frac{\lambda_c}{\lambda_0}=\frac{v_0}{c}$,
where $\lambda_0\sim 1m$.  It is also very curious to observe that
$\lambda_c=\frac{v_0}{c}\lambda_0=\frac{V}{v_0}\lambda_0\Longleftrightarrow
v_0=\sqrt{cV}$,which in fact represents a special intermediate point
 (a kind of {\it aurum point}), namely it represents a geometric average between
$c$ and $V$,where the human scale ($\lambda_0\sim 10^0 m$) is really found as
an intermediate scale. Thus we may write
 $\lambda_c^2=\beta_0^2\lambda_0^2=\alpha_0^2\lambda_0^2$,such that
 $\lambda_c^2=\xi\lambda_0^2\sim 10^{-24}m^2$,where we have
 $\beta_0^2=\alpha_0^2=v_0^2/c^2=V^2/v_0^2=V/c=\xi\sim 10^{-24}$.

 As we already know the total energy $E=m_0c^2\Psi$ and the momentum $\vec P=m_0\vec v\Psi$ at $S_V$,
 we can demonstrate that $E^2=c^2\vec P^2+m_0^2c^4(1-V^2/v^2)$,where $\Psi$ is shown in (71).

 \section{Transformations of space-time and velocity in the presence of the
 ultra-referential $S_V$}

  Let us assume the reference frame $S^{\prime}$ with a speed $v$
 in relation to the ultra-referential $S_V$. To simplify,consider the
 motion only at one spatial dimension,namely $(1+1)D$-space-time with background field
 $S_V$. So we write the following transformations:

  \begin{equation}
 dx^{\prime}=\Psi(dX-\beta_{*}cdt)=\Psi(dX-vdt+Vdt),
  \end{equation}
where $\beta_{*}=\beta\epsilon=\beta(1-\alpha)$,being $\beta=v/c$ and $\alpha=V/v$,so that
$\beta_{*}\rightarrow 0$ for $v\rightarrow V$ or $\alpha\rightarrow 1$.
\footnote{Let us assume the following more general transformations:
$x^{\prime}=\theta\gamma(X-\epsilon_1vt)$ and $t^{\prime}=\theta\gamma(t-\frac{\epsilon_2vX}{c^2})$,
where $\theta$,$\epsilon_1$ and $\epsilon_2$ are factors (functions) to be determined. We hope all these factors
depend on $\alpha$,such that,for $\alpha\rightarrow 0$ ($V\rightarrow 0$),we recover Lorentz transformations
 ($\theta=1$,$\epsilon_1=1$ and $\epsilon_2=1$). By using those transformations to perform
$[c^2t^{\prime 2}-x^{\prime 2}]$,we find the identity: $[c^2t^{\prime 2}-x^{\prime 2}]=
\theta^2\gamma^2[c^2t^2-2\epsilon_1vtX+2\epsilon_2vtX-\epsilon_1^2v^2t^2+\frac{\epsilon_2^2v^2X^2}{c^2}-X^2]$.
 Since the metric tensor is diagonal,the crossed terms must vanish and so we assure that
$\epsilon_1=\epsilon_2=\epsilon$. Due to this fact,the crossed terms ($2\epsilon vtX$) are cancelled between
themselves and finally we obtain $[c^2t^{\prime 2}-x^{\prime 2}]=
 \theta^2\gamma^2(1-\frac{\epsilon^2 v^2}{c^2})[c^2t^2-X^2]$. For $\alpha\rightarrow 0$ ($\epsilon=1$ and
$\theta=1$),we reinstate $[c^2t^{\prime 2}-x^{\prime 2}]=[c^2t^2-x^2]$ of SR. Now we write the following
transformations:$x^{\prime}=\theta\gamma(X-\epsilon vt)\equiv\theta\gamma(X-vt+\delta)$ and
$t^{\prime}=\theta\gamma(t-\frac{\epsilon vX}{c^2})\equiv\theta\gamma(t-\frac{vX}{c^2}+\Delta)$,where
we assume $\delta=\delta(V)$ and $\Delta=\Delta(V)$,such that $\delta=\Delta=0$ for $V\rightarrow 0$,which implies $\epsilon=1$. So from such transformations we extract: $-vt+\delta(V)\equiv-\epsilon vt$ and
$-\frac{vX}{c^2}+\Delta(V)\equiv-\frac{\epsilon vX}{c^2}$,from where we obtain
 $\epsilon=(1-\frac{\delta(V)}{vt})=(1-\frac{c^2\Delta(V)}{vX})$. As $\epsilon$ is a dimensionaless factor,
we immediately conclude that $\delta(V)=Vt$ and $\Delta(V)=\frac{VX}{c^2}$,such that we find
$\epsilon=(1-\frac{V}{v})=(1-\alpha)$. On the other hand,we can determine $\theta$ as follows: $\theta$ is a
function of $\alpha$ ($\theta(\alpha)$),such that $\theta=1$ for $\alpha=0$,which also leads to $\epsilon=1$ in
order to recover Lorentz transformations. So,as $\epsilon$ depends on $\alpha$,we conclude that $\theta$ can
also be expressed in terms of $\epsilon$,namely $\theta=\theta(\epsilon)=\theta[(1-\alpha)]$,where
$\epsilon=(1-\alpha)$. Therefore we can write $\theta=\theta[(1-\alpha)]=[f(\alpha)(1-\alpha)]^k$,where the
exponent $k>0$. The function $f(\alpha)$ and $k$ will be estimated by satisfying the following conditions:
i) as $\theta=1$ for $\alpha=0$ ($V=0$),this implies $f(0)=1$. ii) the function $\theta\gamma=
\frac{[f(\alpha)(1-\alpha)]^k}{(1-\beta^2)^{\frac{1}{2}}}=\frac{[f(\alpha)(1-\alpha)]^k}
{[(1+\beta)(1-\beta)]^{\frac{1}{2}}}$ should have a symmetry behavior,that is to say it goes to zero closer
to $V$ ($\alpha\rightarrow 1$) in the same way it goes to infinite closer to $c$ ($\beta\rightarrow 1$). This
means that the numerator of the function $\theta\gamma$,which depends on $\alpha$ should have the same shape of
its denumerator,which depends on $\beta$. Due to such conditions,we naturally conclude that $k=1/2$ and
$f(\alpha)=(1+\alpha)$,so that $\theta\gamma=
\frac{[(1+\alpha)(1-\alpha)]^{\frac{1}{2}}}{[(1+\beta)(1-\beta)]^{\frac{1}{2}}}=
\frac{(1-\alpha^2)^{\frac{1}{2}}}{(1-\beta^2)^\frac{1}{2}}=\frac{\sqrt{1-V^2/v^2}}{\sqrt{1-v^2/c^2}}=\Psi$,where
$\theta=\sqrt{1-\alpha^2}=\sqrt{1-V^2/v^2}$.}

 \begin{equation}
 dt^{\prime}=\Psi(dt-\frac{\beta_{*}dX}{c})=\Psi(dt-\frac{vdX}{c^2}+\frac{VdX}{c^2}),
  \end{equation}
being $\vec v=v_x{\bf x}$. We have $\Psi=\frac{\sqrt{1-\alpha^2}}{\sqrt{1-\beta^2}}$. If we make
$V\rightarrow 0$ ($\alpha\rightarrow 0$),we recover Lorentz
transformations,where the ultra-referential $S_V$ is eliminated and simply replaced by the Galilean
frame $S$ at rest for the observer.

The transformations shown in (80) and (81) are the direct transformations
from $S_V$ [$X^{\mu}=(X,ict)$] to $S^{\prime}$ [$x^{\prime\nu}=(x^{\prime},ict^{\prime})$],where
we have $x^{\prime\nu}=\Omega^{\nu}_{\mu} X^{\mu}$ ($x^{\prime}=\Omega X$),
so that we obtain the following matrix of transformation:

\begin{equation}
\displaystyle\Omega=
\begin{pmatrix}
\Psi & i\beta (1-\alpha)\Psi \\
-i\beta (1-\alpha)\Psi & \Psi
\end{pmatrix},
\end{equation}
such that $\Omega\rightarrow\ L$ (Lorentz matrix of rotation) for $\alpha\rightarrow 0$
($\Psi\rightarrow\gamma$).

We obtain $det\Omega=\frac{(1-\alpha^2)}{(1-\beta^2)}[1-\beta^2(1-\alpha)^2]$,where $0<det\Omega<1$. Since
$V$ ($S_V$) is unattainable ($v>V)$,this assures that $\alpha=V/v<1$ and therefore the matrix $\Omega$
admits inverse ($det\Omega\neq 0$ $(>0)$). However $\Omega$ is a non-orthogonal matrix
($det\Omega\neq\pm 1$) and so it does not represent a rotation matrix ($det\Omega\neq 1$) in such a space-time
due to the presence of the privileged frame of background field $S_V$ that breaks the invariance of the norm of
4-vector. Such a break occurs strongly closer to $S_V$ because the particle experiments an enormous dislocation (uncertainty) from the origin $O^{\prime}$ of the frame $S^{\prime}$ (see Fig.3). This leads to the strong
inequality $\Delta S^{\prime 2}>>\Delta S^2$,where $\Delta x^{\prime}_v\rightarrow\infty$ for $v\rightarrow V$
(see (55),(56),(57) and (58)). Actually such an effect ($det\Omega\approx 0$ for $\alpha\approx 1$) emerges from such
a new relativistic physics for treating much lower energies at infrared regime (very large wavelengths),where a
new implicit dimension ($\Delta x_5$) ceases to be hidden and then stretches drastically to the infinite closer to
$S_V$ (see (56)). In the limit $S_V$,the ``particle" would loose its identity,by dissolving completely in the
background field ($\Delta x^{\prime}_v=\infty$). So the matrix $\Omega$ would become singular ($det\Omega=0$),
however,as such a limit $V$ is unattainable,this really assures the existence of an inverse matrix for $\Omega$.

 We notice that $det\Omega$ is a function of the speed $v$ with respect to $S_V$. In the approximation for
$v>>V$ ($\alpha\approx 0$),we obtain $det\Omega\approx 1$ and so we practically reinstate the rotation behavior
of Lorentz matrix as a particular regime for higher energies. In this case,we find the particle with a more
determined position ($\Delta x^{\prime}_v\approx 0$),which leads to $\Delta S^{\prime}\approx\Delta S$.
 Alternatively,if we make $V\rightarrow 0$ ($\alpha\rightarrow 0$),we exactly recover $det\Omega=1$ ($\Delta
S^{\prime}=\Delta S$,$\Delta x^{\prime}_v=0$).

The inverse transformations (from $S^{\prime}$ to $S_V$) are

 \begin{equation}
 dX=\Psi^{\prime}(dx^{\prime}+\beta_{*}cdt^{\prime})=\Psi^{\prime}(dx^{\prime}+vdt^{\prime}-Vdt^{\prime}),
  \end{equation}

 \begin{equation}
 dt=\Psi^{\prime}(dt^{\prime}+\frac{\beta_{*}
 dx^{\prime}}{c})=\Psi^{\prime}(dt^{\prime}+\frac{vdx^{\prime}}{c^2}-
\frac{Vdx^{\prime}}{c^2}).
  \end{equation}

In matrix form,we have the inverse transformation $X^{\mu}=\Omega^{\mu}_{\nu} x^{\prime\nu}$
 ($X=\Omega^{-1}x^{\prime}$),so that the inverse matrix is

\begin{equation}
\displaystyle\Omega^{-1}=
\begin{pmatrix}
\Psi^{\prime} & -i\beta (1-\alpha)\Psi^{\prime} \\
 i\beta (1-\alpha)\Psi^{\prime} & \Psi^{\prime}
\end{pmatrix},
\end{equation}
where we can show that $\Psi^{\prime}$=$\Psi^{-1}/[1-\beta^2(1-\alpha)^2]$,so that $\Omega^{-1}\Omega=I$.

 Indeed we have $\Psi^{\prime}\neq\Psi$ and therefore $\Omega^{-1}\neq\Omega^T$. This non-orthogonal aspect of
$\Omega$ has
an important physical implication. In order to understand such an implication,let us consider firstly the
orthogonal (e.g: rotation) aspect of Lorentz matrix in SR. Under SR,we
have $\alpha=0$,so that $\Psi^{\prime}\rightarrow\gamma^{\prime}=\gamma=(1-\beta^2)^{-1/2}$.
 This symmetry ($\gamma^{\prime}=\gamma$, $L^{-1}=L^T$) happens because the Galilean reference
frames allow us to exchange the speed $v$ (of $S^{\prime}$) for $-v$ (of $S$) when we are at rest at
$S^{\prime}$. However,under SSR,since there is no rest at $S^{\prime}$
 (non-Galilean frame),we cannot exchange $v$ (of $S^{\prime}$) for $-v$ (of $S_V$)
due to that asymmetry ($\Psi^{\prime}\neq\Psi$, $\Omega^{-1}\neq\Omega^T$). Due to this fact,
$S_V$ must be covariant,namely $V$ remains invariant for any change of non-Galilean frame. Thus we
can notice that the paradox of twins,which appears due to that symmetry by
exchange of $v$ for $-v$ in SR should be naturally eliminated in SSR,because only the
non-Galilean reference frame $S^{\prime}$ can move with respect to $S_V$ that remains
covariant (invariable for any change of reference frame).

  We have $det\Omega=\Psi^2[1-\beta^2(1-\alpha)^2]\Rightarrow [(det\Omega)\Psi^{-2}]=[1-\beta^2(1-\alpha)^2]$. So
we can alternatively write $\Psi^{\prime}$=$\Psi^{-1}/[1-\beta^2(1-\alpha)^2]=\Psi^{-1}/[(det\Omega)\Psi^{-2}]
=\Psi/det\Omega$. By inserting this result in (85) to replace $\Psi^{\prime}$,we obtain the relationship
between the inverse matrix and the transposed matrix of $\Omega$,namely $\Omega^{-1}=\Omega^T/det\Omega$. Indeed
$\Omega$ is a non-orthogonal matrix,since we have $det\Omega\neq\pm 1$.

 By dividing (80) by (81),we obtain the following speed transformation:

   \begin{equation}
  v_{Rel}=\frac{v^{\prime}-v+V}
{1-\frac{v^{\prime}v}{c^2}+\frac{v^{\prime}V}{c^2}},
   \end{equation}
 where we have considered $v_{Rel}=v_{Relative}\equiv dx^{\prime}/dt^{\prime}$
 and $v^{\prime}\equiv dX/dt$.  $v^{\prime}$ and $v$ are given with
 respect to $S_V$,with $v_{Rel}$ being related between them. Let us consider
 $v^{\prime}>v$. If $V\rightarrow 0$,the transformation (86) recovers the Lorentz
 velocity transformation,where $v^{\prime}$ and $v$ are given in relation to a
 certain Galilean frame $S$ at rest. Since (86) implements the ultra-referential
 $S_V$,the speeds $v^{\prime}$ and $v$ are now given with respect to $S_V$,
 which is covariant (absolute). Such a covariance is verified if we assume that
 $v^{\prime}=v=V$ in (86). Thus,for this case,we obtain $v_{Rel}=``V-V''=V$.
  Let us also consider the following cases:

 {\bf a)} $v^{\prime}=c$ and $v\leq c\Rightarrow v_{Rel}=c$. This
 just verifies the well-known invariance of $c$.

 {\bf b)} if $v^{\prime}>v(=V)\Rightarrow v_{Rel}=``v^{\prime}-V"=v^{\prime}$. For
 example,if $v^{\prime}=2V$ and $v=V$ $\Rightarrow v_{Rel}=``2V-V"=2V$. This
 means that $V$ really has no influence on the speed of the particles. So $V$ works as if
 it were an ``{\it absolute zero of movement}'',being invariant.

 {\bf c)} if $v^{\prime}=v$ $\Rightarrow v_{Rel}=``v-v''$($\neq 0)$
$=\frac{V}{1-\frac{v^2}{c^2}(1-\frac{V}{v})}$. From ({\bf c}) let us
consider two specific cases,namely:

  -$c_1$) assuming $v=V\Rightarrow v_{Rel}=``V-V"=V$ as mentioned before.

  -$c_2$) if $v=c\Rightarrow v_{Rel}=c$,
where we have the interval $V\leq v_{Rel}\leq c$ for $V\leq v\leq c$.

This last case ({\bf c}) shows us in fact that it is impossible to find the
rest for the particle on its own non-Galilean frame $S^{\prime}$,where
$v_{Rel}(v)$ ($\equiv\Delta v(v)$) is an increasing function. However,
if we make $V\rightarrow 0$,then we have $v_{Rel}\equiv\Delta v=0$ and therefore
it would be possible to find the rest for $S^{\prime}$,which becomes a Galilean
reference frame ($v<c$) of SR.

 By dividing (83) by (84),we obtain

 \begin{equation}
  v_{Rel}=\frac{v^{\prime}+v-V}
{1+\frac{v^{\prime}v}{c^2}-\frac{v^{\prime}V}{c^2}}
   \end{equation}

 In (87),if $v^{\prime}=v=V\Rightarrow ``V+V''=V$. Indeed $V$ is
 invariant,working like an {\it absolute zero point} in SSR. If
 $v^{\prime}=c$ and $v\leq c$,this implies $v_{Rel}=c$. For $v^{\prime}>V$
 and considering $v=V$,
 this leads to $v_{Rel}=v^{\prime}$. As a specific example,if $v^{\prime}=2V$
 and assuming $v=V$,we would have $v_{Rel} =``2V+V''=2V$. And if
 $v^{\prime}=v\Rightarrow v_{Rel}=``v+v"=\frac{2v-V}{1+\frac{v^2}{c^2}(1-\frac{V}{v})}$.
 In newtonian regime ($V<<v<<c$),we recover $v_{Rel}=``v+v"=2v$. In relativistic (einsteinian)
regime ($v\rightarrow c$),we reinstate Lorentz transformation for this case ($v^{\prime}=v$),
i.e.,$v_{Rel}=``v+v"=2v/(1+v^2/c^2)$.

 By joining both transformations (86) and (87) into just one,we write the following compact form:

\begin{equation}
  v_{Rel}=\frac{v^{\prime}\mp\epsilon v}
{1\mp\frac{v^{\prime}\epsilon v}{c^2}}=\frac{v^{\prime}\mp v(1-\alpha)}
{1\mp\frac{v^{\prime}v(1-\alpha)}{c^2}}=\frac{v^{\prime}\mp v\pm V}
{1\mp \frac{v^{\prime}v}{c^2}\pm \frac{v^{\prime}V}{c^2}},
\end{equation}
being $\alpha=V/v$ and $\epsilon=(1-\alpha)$. For $\alpha=0$ ($V=0$) or $\epsilon=1$,we recover Lorentz speed
transformations.

 In a more realistic case for motion of the electron in SSR,due to the non-zero minimum limit
of speed $V$ for all directions in the space,actually we should also consider the existence of non-null
transverse components $v_y$ and $v_z$,such that $\vec v_T=v_y{\bf j}+v_z{\bf k}$. So,if we also assume
that such a transverse motion in 2d ($yz$) oscillates in the time ($\vec v_T(t)=v_y(t){\bf j}+v_z(t){\bf k}$)
around $x$,where the particle has a constant longitudinal motion $v=v_x$,we obtain an oscillatory (jittery) motion
for the electron. This so-called {\it zitterbewegung (zbw)} of the electron was introduced by Schroedinger\cite{27}
who proposed the electron spin to be a consequence of a local circulatory motion,constituting {\it zbw} and resulting
from the interference between positive and negative energy solutions of the Dirac equation. Such an issue turned out
to be of renewed interest\cite{28}~\cite{29}. The present work provides naturally a more fundamental vision for {\it
zbw},whose origin is connected to the vacuum energy from the ultra-referential $S_V$,where now gravity also plays an
essential role ($V\propto\sqrt{G}$). We intend to go deeper into such a subject about more general transformations
elsewhere.

 \section{Covariance of the Maxwell wave equation in presence of the ultra-referential $S_V$}

Let us assume a light ray emitted from the frame $S^{\prime}$. Its equation of electrical wave
in this reference frame is

\begin{equation}
\frac{\partial^2\vec E(x^{\prime},t^{\prime})}{\partial x^{\prime 2}}-
\frac{1}{c^2}\frac{\partial^2\vec E(x^{\prime},t^{\prime})}
{\partial t^{\prime 2}}=0
\end{equation}

 As it is already known,when we make the exchange by conjugation on the
spatial and temporal coordinates,we obtain respectively the following
operators: $X\rightarrow\partial/\partial t$ and $t\rightarrow\partial/\partial X$;
also $x^{\prime}\rightarrow\partial/\partial t^{\prime}$ and $t^{\prime}\rightarrow\partial/
\partial x^{\prime}$. Thus the transformations (80) and (81) for such differential operators are

\begin{equation}
\frac{\partial}{\partial t^{\prime}}=\Psi
(\frac{\partial}{\partial t}-\beta c\frac{\partial}{\partial X}+
\xi c\frac{\partial}{\partial X})=
\Psi[\frac{\partial}{\partial t}-\beta
 c(1-\alpha)\frac{\partial}{\partial X})],
\end{equation}

\begin{equation}
\frac{\partial}{\partial x^{\prime}}=\Psi
(\frac{\partial}{\partial X}-\frac{\beta}{c}\frac{\partial}{\partial
 t}+
\frac{\xi}{c}\frac{\partial}{\partial t})=
\Psi[\frac{\partial}{\partial
 X}-\frac{\beta}{c}(1-\alpha)\frac{\partial}{\partial t})],
\end{equation}
where $v=\beta c$, $V=\xi c$ and $\xi=\alpha\beta$,being $\alpha=V/v$.

By squaring (90) and (91),inserting into (89) and after performing the calculations,
we will finally obtain

\begin{equation}
\Psi^2[1-\beta^2(1-\alpha)^2]\left(\frac{\partial^2\vec E}{\partial X^2}
-\frac{1}{c^2}\frac{\partial^2\vec E}{\partial t^2}\right)=
det\Omega\left(\frac{\partial^2\vec E}{\partial X^2}
-\frac{1}{c^2}\frac{\partial^2\vec E}{\partial t^2}\right)=0
\end{equation}

  As the ultra-referential $S_V$ is definitely inaccessible for any particle,we always have
$\alpha<1$ (or $v>V$),which always implies $det\Omega=\Psi^2[1-\beta^2(1-\alpha)^2]>0$. And as we
already have shown in section 6,such a result is in agreement with the fact that we must
have $det\Omega>0$. Therefore this will always assure

  \begin{equation}
 \frac{\partial^2\vec E}{\partial X^2}
-\frac{1}{c^2}\frac{\partial^2\vec E}{\partial t^2}=0
 \end{equation}

By comparing (93) with (89),we verify the covariance of the equation
of the electromagnetic wave propagating in the relativistic ``ether"
 (background field) $S_V$.

\section{Cosmological implications}

\subsection{Energy-momentum tensor in the presence of the ultra-referential-$S_V$}

 Let us write the 4-velocity in the presence of $S_V$,as follows:

   \begin{equation}
 U^{\mu}=\left[\frac{\sqrt{1-\frac{V^2}{v^2}}}{\sqrt{1-\frac{v^2}{c^2}}}~ , ~
\frac{v_{\alpha}\sqrt{1-\frac{V^2}{v^2}}}{c\sqrt{1-\frac{v^2}{c^2}}}\right],
   \end{equation}
where $\mu=0,1,2,3$ and $\alpha=1,2,3$. If $V\rightarrow 0$,we recover the 4-velocity of SR.

The well-known energy-momentum tensor to deal with perfect fluid has the form

   \begin{equation}
  T^{\mu\nu}=(p+\epsilon)U^{\mu}U^{\nu} - pg^{\mu\nu},
   \end{equation}
where now $U^{\mu}$ is given in (94). $p$ represents a pressure and $\epsilon$ an energy density.

   From (94) and (95),by calculating the new component $T^{00}$,we obtain

   \begin{equation}
  T^{00}=\frac{\epsilon(1-\frac{V^2}{v^2})+p(\frac{v^2}{c^2}-\frac{V^2}{v^2})}{(1-\frac{v^2}{c^2})}
   \end{equation}

 If $V\rightarrow 0$,we recover the old component $T^{00}$ of the Relativity theory.

Now,in order to obtain $T^{00}$ in (96) for vacuum limit in the ultra-referential-$S_V$,we perform

  \begin{equation}
 lim_{v\rightarrow V} T^{00}= T^{00}_{vacuum}=\frac{p(\xi^2 -1)}{(1-\xi^2)}= -p,
  \end{equation}
where $\xi=V/c$ (see (42)).

 As we always must have $T^{00}>0$,we have $p<0$ in (97),which implies a negative pressure for vacuum energy
density of the ultra-referential $S_V$. So we verify that a negative pressure emerges naturally from such
new tensor in the limit of $S_V$.

 We can obtain $T^{\mu\nu}_{vacuum}$ by calculating the following limit:

 \begin{equation}
 T^{\mu\nu}_{vacuum}= lim_{v\rightarrow V}T^{\mu\nu}= -pg^{\mu\nu},
 \end{equation}
 where we naturally conclude that $\epsilon=-p$. $T^{\mu\nu}_{vac.}$ is in fact a diagonalized tensor as we
hope to be. So the vacuum-$S_V$,which is inherent to such a space-time works like a {\it sui generis} fluid at
equilibrium and with negative pressure,leading to a cosmological anti-gravity connected to the cosmological
constant.

\subsection{Cosmological constant $\Lambda$}

Let us begin by writing the Einstein equation in the presence of the
cosmological constant $\Lambda$,namely:
\begin{equation}
R_{\mu\nu}-\frac{1}{2}Rg_{\mu\nu}=\frac{8\pi G}{c^2} T_{\mu\nu}+
\Lambda g_{\mu\nu},
\end{equation}
where we think that the anti-gravitational effect due to the vacuum energy
has origin from the last term $\Lambda g_{\mu\nu}$. In the absence of matter
($T_{\mu\nu}=0$),we have
\begin{equation}
R_{\mu\nu}-\frac{1}{2}Rg_{\mu\nu}-\Lambda g_{\mu\nu}=0
\end{equation}

For very large scales of space-time,the presence of the term $\Lambda
g_{\mu\nu}$ is considerable and the accelerated expansion of the
universe is governed by vacuum energy density. So we can relate $\Lambda$ to the
vacuum energy density. To do that,we just use the energy-momentum tensor (95)
(from (94)) given in vacuum limit of the ultra-referential $S_V$
 (see (98)). Thus we can rewrite equation (100) in its equivalent form for the
energy-momentum tensor given in the limit of vacuum-$S_V$,as follows:
\begin{equation}
R_{\mu\nu}-\frac{1}{2}Rg_{\mu\nu}-\frac{8\pi G}{c^2} T_{\mu\nu}^{vac.}=0,
\end{equation}
where $T_{\mu\nu}^{vac.}=lim_{v\rightarrow V}T_{\mu\nu}= -pg_{\mu\nu}$
(see (98)). And as $p=-\epsilon=-\epsilon_{vac.}=-\rho_{(\Lambda)}$ ($p=w\epsilon$ with $w=-1$),
we write (101) in the following way:
\begin{equation}
R_{\mu\nu}-\frac{1}{2}Rg_{\mu\nu}-\frac{8\pi
G}{c^2}\rho_{(\Lambda)}g_{\mu\nu}=0
\end{equation}

Finally,by comparing (102) with (100),we obtain
\begin{equation}
\rho_{(\Lambda)}=\frac{\Lambda c^2}{8\pi G},
\end{equation}
which gives the direct relationship between cosmological constant $\Lambda$
and vacuum energy density $\rho_{(\Lambda)}$.

Our aim is to estimate $\Lambda$ and $\rho_{(\Lambda)}$ by using the idea of such
a universal minimum speed $V$ and its influence on
gravitation at very large scales of length. In order to study such an
influence,let us firstly start from the well-known simple model of a massive particle that
escapes from a classical gravitational potential $\phi$,where its total
relativistic energy for an escape velocity $v$ is due to the
presence of such a potential $\phi$,namely
$E=mc^2(1-v^2/c^2)^{-1/2}\equiv mc^2(1+\phi/c^2)$. Here the interval of velocity $0\leq v<c$
is associated with the interval of potential $0\leq\phi<\infty$,where we
stipulate $\phi> 0$ to be attractive potential. Now it is very
important to notice that the breakdown of Lorentz symmetry due to
$S_V$ of background field has origin in a non-classical (non-local) aspect
of gravitation that leads to a repulsive gravitational potential
($\phi<0$) for very large distances (cosmological anti-gravity). In order to see such a modified aspect
of gravitation\cite{30},let us consider the total energy of the particle with respect to $S_V$,shown in
(72),namely:

\begin{equation}
E= m_0c^2\frac{\sqrt{1-\frac{V^2}{v^2}}}{\sqrt{1-\frac{v^2}{c^2}}}\equiv
m_0c^2(1+\phi/c^2),
\end{equation}
from where we obtain
\begin{equation}
\phi\equiv c^2\left[\frac{\sqrt{1-\frac{V^2}{v^2}}}{\sqrt{1-\frac{v^2}{c^2}}}-1\right]
\end{equation}

From (105),we observe two regimes of gravitational potential,namely:
\begin{equation}
\phi= \left\{
\begin{array}{ll}
\phi_R:&\mbox{$-c^2<\phi\leq 0$ for $V(=\xi c)< v\leq v_0$},\\\\
  \phi_A:&\mbox{$0\leq\phi<\infty$ for $v_0(=\sqrt{\xi}c)\leq v< c$}.
\end{array}
\right.
\end{equation}
$\phi_A$ and $\phi_R$ are respectively the attractive (classical) and repulsive (non-classical)
potentials. We observe that the strongest repulsive potential is
$\phi=-c^2$,which is associated with a vacuum energy for the
ultra-referential $S_V$ of the universe as a whole (consider $v=V$ in (105)). Therefore
such most negative potential is related to the cosmological constant (see (97)),and so we write:

\begin{equation}
\phi_{\Lambda}=\phi(V)=-c^2
\end{equation}

  The negative potential above depends directly on $\Lambda$,namely
$\phi_{\Lambda}=\phi(\Lambda)=\phi(V)=-c^2$. To show that,let us
consider a simple model of spherical universe with a radius $R_u$,being filled by
a uniform vacuum energy density $\rho_{(\Lambda)}$,so that the total vacuum energy inside the sphere is
$E_{\Lambda}=\rho_{(\Lambda)}V_u=-pV_u=M_{\Lambda}c^2$. $V_u$ is its volume and $M_{\Lambda}$ is the total
dark mass associated with the dark energy for $\Lambda$ ($w=-1$). Therefore the repulsive gravitational
potential on the surface of such a sphere is
\begin{equation}
\phi_{\Lambda}=-\frac{GM_{\Lambda}}{R_u}=-\frac{G\rho_{(\Lambda)}V_u}{R_uc^2}=\frac{4\pi GpR_u^2}{3c^2}
\end{equation}

 By introducing (103) into (108),we find
\begin{equation}
\phi_{\Lambda}=\phi(\Lambda)=-\frac{\Lambda R_u^2}{6}
\end{equation}

Finally,by comparing (109) with (107),we obtain

\begin{equation}
\Lambda=\frac{6c^2}{R_u^2},
\end{equation}
where $\Lambda S_u=24\pi c^2$,being $S_u=4\pi R_u^2$.

And also by comparing (108) with (107),we have
\begin{equation}
\rho_{(\Lambda)}=-p=\frac{3c^4}{4\pi G R_u^2},
\end{equation}
where $\rho_{(\Lambda)} S_u=3c^4/G$. (111) and (110) satisfy (103).

 $\Lambda$ (eq. 110) is a kind of {\it cosmological scalar field}, extending the old concept of
Einstein about the cosmological constant for stationary universe. From (110),by considering the Hubble radius,with
$R_{u}=R_{H_0}\sim 10^{26}m$,we obtain $\Lambda=\Lambda_0\sim (10^{17}m^2s^{-2}/10^{52}m^2)\sim 10^{-35}s^{-2}$.
To be more accurate,we know the age of the universe $T_0=13.7$ Gyr,being $R_{H_0}=cT_0\approx
1.3\times 10^{26}m$,which leads to $\Lambda_0\approx 3\times 10^{-35}s^{-2}$. This result is very close to
the observational results\cite{31}\cite{32}\cite{33}\cite{34}\cite{35}. The tiny vacuum energy
density\cite{36}\cite{37} shown in (111) for $R_{H_0}$ is $\rho_{(\Lambda_0)}\approx
2\times 10^{-29}g/cm^{3}$,which is also in agreement with observations. For scale of the Planck length,where
$R_{u}=l_P=(G\hbar/c^3)^{1/2}$,from (110) we find $\Lambda=\Lambda_P=6c^5/G\hbar\sim 10^{87}s^{-2}$, and from (111)
$\rho_{(\Lambda)}=\rho_{(\Lambda_P)}=T^{00}_{vac.P}=\Lambda_P c^2/8\pi G=3c^7/4\pi G^2\hbar\sim 10^{113}J/m^3
(=3c^4/4\pi l_P^2G\sim 10^{43}kgf/S_P\sim 10^{108}atm\sim 10^{93}g/cm^3)$. So just at that past time,$\Lambda_P$ or
$\rho_{(\Lambda_P)}$ played the role of an inflationary vacuum field with 122 orders of magnitude\cite{38} beyond of
those ones ($\Lambda_0$ and $\rho_{(\Lambda_0)}$) for the present time.

It must be emphasized that our assumption for obtaining the tiny value of $\Lambda$ starts from new fundamental principles
in the space-time. So it does not depend on detailed adjustments with cosmological models.

The study of competition between gravity and anti-gravity ($\Lambda$) during the expansion of the universe
will be treated elsewhere.

\section{Conclusions and prospects}

  We have introduced a space-time with symmetry,so that $V<v\leq c$,
 where $V$ is an inferior and unattainable limit of speed
 associated with a privileged inertial reference frame of universal background field.
  So we have essentially concluded that the space-time structure
 where gravity is coupled to electromagnetism at quantum level naturally
 contains the fundamental ingredients for comprehension of the quantum
 uncertainties through that mentioned symmetry ($V<v\leq c$),where
gravity plays a crucial role due to the minimum velocity $V(\propto G^{1/2})$
related to the minimum length (Planck scale) of DSR\cite{20}\cite{21}\cite{22}\cite{23}
\cite{24}\cite{25} by Magueijo,Smolin,Camelia,et al.

We have studied the cosmological implications of $S_V$,by estimating the tiny values of the
vacuum energy density ($\rho_(\Lambda)=10^{-29}g/cm^3$) and the current cosmological
constant ($\Lambda\sim 10^{-35}s^{-2}$),which are still not well understood by quantum
field theories for quantum vacuum\cite{38},because such theories foresee a very high
value for $\Lambda$,whereas,on the other hand,exact supersymmetric theories foresee an exact null
value for it,which also does not agree with Reality.

 The present theory has various implications which shall be
investigated in coming articles. A new transformation group for such a space-time
will be explored in details. We will propose the development of a new relativistic dynamics,
where the energy of vacuum (ultra-referential $S_V$) plays a crucial role for understanding
the origin of the inertia,including the problem of mass anisotropy.

Another relevant investigation is with respect to the problem of the
absolute zero temperature in thermodynamics of a gas. We intend to make a connection between
the 3rd. law of Thermodynamics and the new dynamics,through a relationship
between the absolute zero temperature ($T=0K$) and the minimum average speed
 ($\left<v\right>_N=V$) for $N$ particles. Since $T=0K$ is
thermodynamically unattainable,this is due to the impossibility of reaching
 $\left<v\right>_N=V$ from the new dynamics standpoint. This leads still to other important
 implications,such as for example,Einstein-Bose condensate and the problem of
 the high refraction index of ultracold gases,where we intend to estimate that
 the speed of light would approach to $V$ inside the condensate medium for $T\rightarrow
 0K$. So the maximum refraction index would be $n_{max}=c/v_{min}=
 c/V=\xi^{-1}=\sigma\sim 10^{23}$ to be shown elsewhere. Thus we will be in a condition to
 propose an experimental manner of making an extrapolation in order to obtain
 $v_{lightMin.}=c^{\prime}_{min}\rightarrow V$ for $T\rightarrow 0K$,through a
 mathematical function obtained by the theory applied to ultracold systems.

 In sum,we begin to open up a new fundamental research field for various areas of Physics,by
including condensed matter,quantum field theories,cosmology (dark energy and cosmological constant)
and specially a new exploration for quantum gravity at very low energies (very large wavelengths).\\

{\noindent\bf  Acknowledgedments}

  This research has been supported by Prof.J.A.Helayel-Neto,from
 {\it Brasilian Center of Physical Research (CBPF-Rio de Janeiro-Br)}. I am
 profoundly grateful to my friends Prof.Carlos Magno Leiras and  Prof.Wladimir Guglinski
 for very interesting colloquies during the last 15 years. I am also very grateful to Prof.C.Joel Franco,
 Prof.Herman J.M.Cuesta, R.Turcati and J.Morais {\it (ICRA:Inst.of Cosmology,Relativity,Astrophysics-CBPF)}.

\end{document}